# Calculation of molecular *g*-tensors by sampling spin orientations of generalised Hartree-Fock states

Shadan Ghassemi Tabrizi,[1,a,*] R. Rodríguez-Guzmán,[2] and Carlos A. Jiménez-Hoyos[1,†]

[1]*Department of Chemistry, Wesleyan University, Middletown, CT 06459, USA*

[2]*Department of Applied Physics I, University of Sevilla, Sevilla, E-41011, Spain*

[a]*Present address: Department of Chemistry, University of Potsdam, Karl-Liebknecht-Str. 24-25, D-14476, Potsdam-Golm, Germany*

*[*]shadan_ghassemi@yahoo.com*

*[†]cjimenezhoyo@wesleyan.edu*

**Abstract.** The variational inclusion of spin-orbit coupling in self-consistent field (SCF) calculations requires a generalised two-component framework, which permits the single-determinant wave function to completely break spin symmetry. The individual components of the molecular *g*-tensor are commonly obtained from separate SCF solutions that align the magnetic moment along one of the three principal tensor axes. However, this strategy raises the question if energy differences between solutions are relevant, or how convergence is achieved if the principal axis system is not determined by molecular symmetry. The present work resolves these issues by a simple two-step procedure akin to the generator coordinate method (GCM). First, a few generalised Hartree Fock (GHF) solutions are converged, applying, where needed, a constraint to the orientation of the magnetic-moment or spin vector. Then, superpositions of GHF determinants are formed through non-orthogonal configuration interaction. This procedure yields a Kramers doublet for the calculation of the complete *g*-tensor. Alternatively, for systems with weak spin-orbit effects, diagonalisation in a basis spanned by spin rotations of a single GHF determinant affords qualitatively correct *g*-tensors by eliminating errors related to spin contamination. For small first-row molecules, these approaches are evaluated against experimental data and full configuration interaction results. It is further demonstrated for two systems (a fictitious tetrahedral $CH_4^+$ species, and a $CuF_4^{2-}$ complex) that a GCM strategy, in contrast to alternative mean-field methods, can correctly describe the spin-orbit splitting of orbitally-degenerate ground states, which causes large *g*-shifts and may lead to negative *g*-values.

# 1. Introduction

The *g*-tensor **g** is a fundamental quantity in the phenomenological description of electron paramagnetic resonance (EPR) spectra of molecules with unpaired electrons. The spin Hamiltonian $\hat{H}_S$,

$$\hat{H}_S = \mu_B \mathbf{B} \cdot \mathbf{g} \cdot \tilde{\mathbf{S}} ,\qquad(1)$$

defines the magnetic-field (**B**) dependent Zeeman splitting of a Kramers doublet, where the latter is represented by an $\tilde{S} = \frac{1}{2}$ pseudospin $\tilde{\mathbf{S}}$ [1]. In molecules with light atoms (through the 3*d* series) the pseudospin is usually perturbatively related to a true electronic spin $S = \frac{1}{2}$ in the absence of spin-orbit coupling (SOC). However, the use of Eq. (1) is not restricted to such cases [1,2].

To help interpret spectra in terms of electronic structure, quantum-chemical calculations of *g*-tensors must consider SOC as the main cause of deviations from the isotropic free-electron *g*-value, $g_e \approx 2.0023193$. From a multitude of computational methods (see, e.g., Refs. [3–5] for additional literature), we mention the coupled-perturbed treatment of SOC by Hartree Fock (HF) theory or – more commonly – Kohn-Sham (KS) density functional theory (DFT) [6,7]; generally more accurate *ab initio* methods include coupled-cluster theory [8], complete active space self-consistent field (CASSCF) [9,10], and multi-reference configuration interaction [11,12]. In the spin-orbit state-interaction (SOSI) procedure, a few selected nonrelativistic (or scalar relativistic) CASSCF solutions form a space for the subsequent diagonalisation of a full Hamiltonian that includes SOC. The *g*-tensor is calculated from the resulting Kramers doublet (see Eq. (4) in the Theory section below) [9,10]. In case of strong mixing between many nonrelativistic states in systems featuring heavy atoms, SOC should be treated in a one-step procedure to calculate the Kramers doublet [13]. Dynamic correlation effects are often of relatively minor importance for *g*-tensors [9], but may be captured by multi-reference perturbation theory [5,9,14]. DMRG-CASSCF can handle larger active spaces, thereby enabling *g*-tensor calculations in rather large strongly correlated systems [4,15], e.g., multinuclear transition-metal clusters [4].

Despite such progress, simple and reliable low-scaling approaches, preferably with a mean-field cost and applicable to a wide range of systems, are still of high interest. Specifically two- or four-component (2c, 4c) HF or DFT approaches implicitly take into account higher-order spin-orbit effects, important for *g*-tensors in 4*d*, 5*d* or actinide complexes, by treating SOC self-consistently [3,16–19]. A common strategy is to seek SCF solutions for different orientations of the magnetic moment and to derive each *g*-tensor component from the respective expectation value $\langle \hat{\boldsymbol{\mu}} \rangle$ [20]. Three solutions with $\langle \hat{\boldsymbol{\mu}} \rangle$ aligned along the principal axes **x**, **y** and **z**, related to elements of the molecular point-group, afford the principal *g*-components in rhombic molecules [20]. This approach is hence called 3SCF [3]. However, for lower molecular symmetry, up to six different SCF solutions are needed to determine all components,[1] possibly with $\langle \hat{\boldsymbol{\mu}} \rangle$ pointing along the Cartesian axes or bisectors [20], but SCF-convergence onto a desired $\langle \hat{\boldsymbol{\mu}} \rangle$ orientation from some initial guess may not be guaranteed.

In addition, the somewhat unsatisfactory fact that each *g*-component is determined from a separate SCF solution has raised conceptual questions [21]. In the presence of SOC, the different solutions are in general not symmetry related and therefore not degenerate. The energy anisotropy with respect to different orientations of $\langle \hat{\boldsymbol{\mu}} \rangle$ or $\langle \hat{\mathbf{S}} \rangle$ is thus expected (and was quantitatively explored for a few $\tilde{S} = \frac{1}{2}$ systems in Ref. [22]) and does not demand a vaguely hypothesised [21] revision of the foundations of EPR spectroscopy. It is however not clear if or how the anisotropy affects *g*-tensor predictions from mean-field calculations. Note that a similar conceptual issue does not arise in other approaches [9,10,23]. Specifically, the calculation of the physically relevant quantities [2] $\mathbf{G} = \mathbf{g}\mathbf{g}^T$ (Eq. (4) below) and $\det(\mathbf{g})$ is unequivocal when based on a qualitatively correct CASSCF-type wave function or, more generally, a basis-set exact full configuration interaction (FCI) wave function. In other words, the calculation of the field-dependent splitting, determined by **G** (in the limit $\mathbf{B} \to 0$), and the prediction of the sense of precession of the magnetic moment around the field direction, determined by the sign of $\det(\mathbf{g})$, are unambiguous.

This work proposes a mean-field approach that is closely related to 3SCF, but conceptually more satisfactory and of broader applicability. It is based on a manifold of constrained HF solutions with different orientations of $\langle \hat{\boldsymbol{\mu}} \rangle$ (or $\langle \hat{\mathbf{S}} \rangle$, or $\langle \hat{\mathbf{L}} \rangle$). We discuss straightforward $\langle \hat{\boldsymbol{\mu}} \rangle$-constrained optimisation based on a Thouless parametrisation of GHF-type Slater determinants along with an optimisation library that can handle nonlinear constraints. Diagonalisation in this manifold yields a single Kramers doublet that determines all components of a qualitatively correct *g*-tensor. Alternatively, a suitable manifold can be spanned by spin rotations of a single

---
[1] In low-symmetry systems, the *g*-tensor may in principle have nine independent components. However, it is always possible to eliminate the three antisymmetric components by pseudospin rotations, see Ref. [1].

GHF solution. Where required, orbital degeneracies can be treated by considering orientational manifolds associated with several symmetry-broken configurations.

## 2. Theory and Computations

The spin Hamiltonian of Eq. (1) generically describes a tensorial linear field-induced level splitting (the lifting of Kramers degeneracy). A connection to electronic-structure theory is established by identifying the electronic Zeeman term, Eq. (2), as the cause of the splitting,

$$\hat{H}_{\text{Zee}} = -\mathbf{B} \cdot \hat{\boldsymbol{\mu}} \; . \tag{2}$$

Let $\boldsymbol{\mu} = (\boldsymbol{\mu}_x, \boldsymbol{\mu}_y, \boldsymbol{\mu}_z)$ denote the $2 \times 2$ matrices of the three Cartesian components of the electronic magnetic moment $\hat{\boldsymbol{\mu}} = \mu_B(g_e \hat{\mathbf{S}} + \hat{\mathbf{L}})$ in the Kramers-pair basis of time-reversal conjugate electronic states, $|\Phi\rangle$ and $|\bar{\Phi}\rangle \equiv \hat{\Theta}|\Phi\rangle$. Up to a constant factor, the components of $\mathbf{g}$ are the coefficients in an expansion of $\boldsymbol{\mu} = (\boldsymbol{\mu}_x, \boldsymbol{\mu}_y, \boldsymbol{\mu}_z)$ in terms of Pauli spin matrices [2]. The specific component values depend on the definition of pseudospin functions, meaning the establishment of a one-to-one correspondence between pseudospin states $\{|\uparrow\rangle, |\downarrow\rangle\}$ and electronic states $\{|\Phi\rangle, |\bar{\Phi}\rangle\}$ [1,2], but the symmetric Abragam-Bleaney tensor [2,24] $\mathbf{G} = \mathbf{g}\mathbf{g}^T$ is an invariant [2] that defines the splitting $\Delta E$, Eq. (3),

$$\Delta E = \mu_B \sqrt{\sum_{m,n} B_m B_n G_{mn}} \; , \tag{3}$$

resulting from the diagonalisation of $\hat{H}_{\text{Zee}}$ in the basis $\{|\Phi\rangle, |\bar{\Phi}\rangle\}$. The elements $G_{mn}$ ($m,n = x,y,z$) of $\mathbf{G}$ were derived by Gerloch and McMeeking [24],

$$G_{mn} = 2 \sum_{v=\Phi,\bar{\Phi}} \sum_{w=\Phi,\bar{\Phi}} \langle v|\hat{\mu}_m|w\rangle \langle w|\hat{\mu}_n|v\rangle \; . \tag{4}$$

Diagonalisation of $\mathbf{G}$ amounts to a rotation to its principal-axis system. The square roots of the eigenvalues yield the principal values of the diagonal tensor $\mathbf{g} = \text{diag}(g_x, g_y, g_z)$, $g_n = \sqrt{G_n}$, although, to be precise, each individual value is only defined up to a sign, $g_n = \pm\sqrt{G_n}$ (see below). We report shifts $\Delta g_n = g_n - g_e$ in ppm ($10^{-6}$). The outlined approach, advocated by Bolvin [9], who also cited earlier applications of Eq. (4), is a standard recipe in conjunction with SOSI between nonrelativistic (or scalar relativistic) CASSCF solutions [10].

The quantity det(**g**) is another invariant (unchanged by arbitrary unitary transformations among pseudospin functions). Its sign corresponds to the sign of the product $g_x g_y g_z$ [1,2], which defines the sense of precession of $\langle \hat{\boldsymbol{\mu}} \rangle$ and is relevant for the absorption of circularly polarised radiation [2]. Although the signs of the individual principal components are arguably not meaningful for the prediction of any property, they may still be fixed by the following simple procedure: increase the SOC strength stepwise from zero, where pseudospin is equivalent to true spin (thus, $\mathbf{g} = g_e$), until a realistic SOC strength is reached, while successively updating the pseudospin functions by a Bloch [25]/des Cloizeaux [26] perturbative connection to the respective functions at the previous step. This in essence describes the adiabatic-connection strategy of Chibotaru [1]; see also Ref. [27] for simple applications in a different context, including $\tilde{S} > \frac{1}{2}$ systems. In other cases, high molecular symmetry anchors the adiabatic connection and fixes the signs of principal $g$-values [1,28]. However, we presently do not consider the formal problem of pseudospin definition. In our test set of light molecules, SOC effects are weak, keeping pseudospin closely related to true spin. Hence, all components of **g** are positive. Besides, rhombic symmetry determines the principal axis system.

On the other hand, two artificial tetrahedral model systems ($CH_4^+$ and $CuF_4^{2-}$) display orbitally degenerate ground states and serve to demonstrate that a GCM approach (explained below) can also describe the first-order SOC-induced level-splitting of orbitally degenerate states, where pseudospin is not anymore related to true spin and negative $g$-values may ensue. In these cases, tetrahedral symmetry enforces isotropy, $g_x = g_y = g_z$ [28]. The sign of $g_x g_y g_z$ corresponds to the sign of the quotient $g_x g_z / g_y$,

$$\frac{g_x g_z}{g_y} = -\frac{i}{\mu_B} \frac{(\boldsymbol{\mu}_x \boldsymbol{\mu}_z - \boldsymbol{\mu}_z \boldsymbol{\mu}_x)_{ab}}{(\boldsymbol{\mu}_y)_{ab}}, \qquad (5)$$

where $a,b = 1,2$ is an arbitrary index combination for a non-zero element $(\boldsymbol{\mu}_y)_{ab}$ of the $2 \times 2$ matrix $\boldsymbol{\mu}_y$. Eq. (5), derived in Ref. [23], adheres to the usual Condon-Shortley phase convention.

The 3SCF procedure outlined in the Introduction was first applied in the frame of GHF [20] (including one- and two-electron spin-orbit terms) and later also in quasi-relativistic 2c-DFT [19,29,30] or 4c-Dirac-KS calculations [17]. Our present focus is on GHF, which breaks spin symmetry by independently expanding the ↑ and ↓ spin components of molecular orbitals $\psi_i$ in terms of spatial basis functions $\phi_i$,

$$\psi_i = \begin{pmatrix} \sum_j C_{ji}^\uparrow \phi_j \\ \sum_k C_{ki}^\downarrow \phi_k \end{pmatrix} , \quad (6)$$

with generally complex coefficients $C_{ji}^\uparrow$ and $C_{ki}^\downarrow$. GHF solutions occur only in specific situations in nonrelativistic systems [31], but are the rule when SOC is included [32].

Besides the one-electron SOC term, the quasi-relativistic Breit-Pauli Hamiltonian contains spin-same-orbit (SSO) and spin-other-orbit (SOO) two-electron terms. The burden on memory and computation times [33] incurred by SSO and SOO integrals is often alleviated by a mean-field approximation [34]. However, for ultimate simplicity, we account for two-electron SOC only implicitly through effective nuclear charges $Z_{eff}(K)$ of Koseki et al. [35] in the one-electron SOC term, which we deem adequate, at least when aiming for semi-quantitative accuracy.

$$\hat{H}_{SOC} = \frac{\alpha^2}{2} \sum_{i,K} Z_{eff}(K) r_{iK}^{-3} (\hat{\mathbf{r}}_{iK} \times \hat{\mathbf{p}}_i) \cdot \hat{\mathbf{s}}_i \quad (7)$$

In Eq. (7), $\alpha$ is the fine-structure constant, and $i$ and $K$ are electron and nuclear indices, respectively. $\hat{H}_{SOC}$ is added to the nonrelativistic electronic Hamiltonian $\hat{H}_0$, $\hat{H} = \hat{H}_0 + \hat{H}_{SOC}$.

If the principal axis system is known, 3SCF relies on solutions $|\Phi_\mathbf{n}\rangle$ that align the magnetic moment along the principal unit vectors $\mathbf{n} = \mathbf{x}, \mathbf{y}, \mathbf{z}$,

$$\frac{\langle \Phi_\mathbf{n} | \hat{\boldsymbol{\mu}} | \Phi_\mathbf{n} \rangle}{|\langle \Phi_\mathbf{n} | \hat{\boldsymbol{\mu}} | \Phi_\mathbf{n} \rangle|} = \mathbf{n} . \quad (8)$$

In rhombic systems, $|\Phi_\mathbf{n}\rangle$ will usually have self-consistent symmetry that enforces perfect alignment between magnetic moment and spin, that is, $\langle \Phi_\mathbf{n} | \hat{\mathbf{S}} | \Phi_\mathbf{n} \rangle \propto \mathbf{n}$ (see Results section). An "effective" $g$-value for a given direction is equated to the component of $\hat{\boldsymbol{\mu}}$ along $\mathbf{n}$,

$$\sqrt{\mathbf{n}^T \cdot \mathbf{G} \cdot \mathbf{n}} = |\langle \Phi_\mathbf{n} | \mathbf{n} \cdot \hat{\boldsymbol{\mu}} | \Phi_\mathbf{n} \rangle| . \quad (9)$$

Eq. (9) may also be applied when $\mathbf{n}$ is not a principal tensor axis, although such cases have rarely been treated [20]. If $\mathbf{n}$ is indeed a principal axis, $g_n$ (assumed positive; $n = x, y, z$) is obtained from Eq. (10),

$$g_n = \left|\langle \Phi_\mathbf{n} | \mathbf{n} \cdot \hat{\boldsymbol{\mu}} | \Phi_\mathbf{n} \rangle \right| . \tag{10}$$

As an alternative to 3SCF, Cherry et al. [3] applied Eq. (4) based on one selected SCF solution, e.g., $\{|\Phi_\mathbf{z}\rangle, |\bar{\Phi}_\mathbf{z}\rangle\}$. They explained that the perpendicular components ($g_x$ and $g_y$ in this example) are strongly affected by spin contamination: for the UHF solution ($\hat{H}_{SOC} = 0$), $g_z = g_e$, but $g_x = g_y \neq g_e$. While a quasi-restricted ansatz [36] avoids this complication by reverting to ROHF for $\hat{H}_{SOC} = 0$, it misses important spin-polarisation effects [3,19]. Thus, an *a posteriori* correction was proposed for spin-contamination errors in *g*-tensors from completely unrestricted mean-field states (here corresponding to GHF) [3]. However, such corrections become ill-defined unless SOC is relatively weak. Besides, the resulting *g*-tensor is at least moderately dependent on the choice of the specific GHF solution [3]. Lastly, this procedure does not attempt to generate a qualitatively correct Kramers pair.

In the present work, we use a generator coordinate method (GCM) ansatz, Eq. (11), for the ground state,

$$|\Psi\rangle = \int d\Omega\, |\Phi_\Omega\rangle f(\Omega) , \tag{11}$$

where we create a superposition of constrained HF solutions $|\Phi_\Omega\rangle$ with different spherical orientations $\Omega = (\vartheta, \phi)$ of $\langle \hat{\boldsymbol{\mu}} \rangle$ (or, alternatively, $\langle \hat{\mathbf{S}} \rangle$ or $\langle \hat{\mathbf{L}} \rangle$). The linear coefficients $f(\Omega)$ are obtained using the variational principle by solution to the Griffin-Hill-Wheeler (GHW) equation,

$$\int d\Omega [H(\Omega, \Omega') - ES(\Omega, \Omega')] f(\Omega') = 0 , \tag{12}$$

with $H(\Omega, \Omega') = \langle \Phi_\Omega | \hat{H} | \Phi_{\Omega'} \rangle$ and $S(\Omega, \Omega') = \langle \Phi_\Omega | \Phi_{\Omega'} \rangle$.

In practical calculations, Eq. (12) is discretised by a numerical sampling of the three-dimensional set of $\langle \hat{\boldsymbol{\mu}} \rangle$ orientations. The GHW equation becomes then equivalent to a non-orthogonal configuration interaction (NOCI) problem, which can be solved as a standard generalised eigenvalue problem for the Kramers doublet of interest. We emphasise, however, that the GCM ansatz of Eq. (12) implies the existence of a smooth, square-integrable function $f(\Omega)$. Note that, in the absence of SOC, the GCM ansatz of Eq. (11) by sampling all possible spin orientations of a UHF state becomes equivalent to a spin-projection ansatz (applied in a projection-after-variation framework, PAV). The spin-projected HF state has, by construction, no spin contamination and correctly yields a vanishing *g*-shift. The latter fact prompted us to also study an alternative to converging GHF for different $\langle \hat{\boldsymbol{\mu}} \rangle$ orientations, by spanning a basis from spin rotations of a single GHF solution to eliminate spin-contamination errors (details are provided in the Results section below).

The evaluation of matrix elements $H(\Omega, \Omega')$ between non-orthogonal Slater determinants [37] is explained in detail in Ref. [38] and is not repeated here. Kramers partners are orthogonal, $\langle \Phi | \bar{\Phi} \rangle = 0$, and do not couple, $\langle \Phi | \hat{H} | \bar{\Phi} \rangle = 0$ (matrix elements of time-even operators, $\hat{H} = \hat{\Theta}^{-1} \hat{H} \hat{\Theta}$, are zero between time-reversal conjugates, if $\hat{\Theta}^2 = -1$ [2]). It is thus straightforward to set up the NOCI problem, Eq. (13), in terms of Hamiltonian and overlap matrices, **H** and **S**,

$$\mathbf{HC} = \mathbf{SCE} \ , \tag{13}$$

where **E** is a diagonal energy matrix. For some of the larger spaces that we consider, the NOCI basis displays near linear dependencies, which we remove [39] based on a threshold of $10^{-8}$ for the eigenvalues of **S**. Two columns $\{\mathbf{v}, \bar{\mathbf{v}}\}$ of the solution matrix **C** represent the Kramers doublet $\{|\Phi\rangle, |\bar{\Phi}\rangle\}$ of interest (the doubly-degenerate ground state, unless noted otherwise) and are used to calculate **G** from Eq. (4), based on NOCI-basis representations $\boldsymbol{\mu}_n$ ($n = x, y, z$) of the magnetic moment. Due to orthogonality, $\langle \Phi | \bar{\Phi} \rangle = 0$, the matrix element $\langle \Phi | \hat{w} | \bar{\Phi} \rangle$ of a one-electron operator $\hat{w}$ (here, $\hat{w} = \mathbf{n} \cdot \hat{\boldsymbol{\mu}}$) requires special attention. In Eq. (14), adopted from Ref. [3],

$$\langle \Phi | \hat{w} | \bar{\Phi} \rangle = \frac{1}{N} \sum_{k=1}^{N} \det(\mathbf{w}_k) \ , \tag{14}$$

$N$ is the number of electrons and $\mathbf{w}_k$ is an $N \times N$ matrix with elements

$$(\mathbf{w}_k)_{ij} = \begin{cases} w_{ij}, & \text{if } k = i \\ s_{ij}, & \text{else} \end{cases} , \tag{15}$$

and $w_{ij} = \langle \psi_i | \hat{w} | \bar{\psi}_j \rangle$ is a matrix element between occupied molecular orbitals in $|\Phi\rangle$ and $|\bar{\Phi}\rangle$, and $s_{ij} = \langle \psi_i | \bar{\psi}_j \rangle$ is the respective overlap.

One- and two-electron integrals are read from the `Gaussian` program [40] to perform GHF and GCM calculations in our set of `Matlab` programs. Gradient-based constrained GHF optimisation is detailed in Appendix A. Even though we generally do not use the common self-consistent procedure of successively building and diagonalising a Fock matrix, we may still use the term "SCF solution" as a synonym for a GHF state. In our small test systems, the centers of nuclear or electronic charge, or spin density, are at similar locations. We place the coordinate origin as the reference point for $\hat{\mathbf{L}}$, which enters the definition of $\hat{\boldsymbol{\mu}} = \mu_B(g_e\hat{\mathbf{S}} + \hat{\mathbf{L}})$, at the center of nuclear charge, as recommended in a previous work that found only a very modest gauge dependence [9].

We first apply our GCM strategy to a small set of light main-group $S = \frac{1}{2}$ radicals ($CO^+$, CN, MgF, $NO_2$, $NF_2^-$, $CO_2^-$, $O_3$, $H_2O^+$, $H_2CO^+$, all structures taken from Ref. [12], the cc-pVTZ [41,42] basis set is used) that were already studied by Jayatilaka upon introducing the 3SCF method [20]. To provide an unambiguous benchmark, we also compute $g$-tensors from FCI for a few systems, using another in-house program. As the feasibility of our FCI calculations rests on basis truncation, we work with unrestricted natural orbitals. UNOs are the eigenfunctions (with non-zero eigenvalues) of the UHF ($\hat{H}_{SOC} = 0$) charge-density matrix. Their number equals the number of electrons, e.g., $N = 9$ in $H_2O^+$. We however restrain the lowest-energy orbitals that correspond to $1s$ shells (excluding $1s$ orbitals on H) to be doubly occupied in the UHF calculations that determine UNOs, and subsequently perform FCI in this frozen-core UNO basis (FC-UNO) to further reduce the number of configurations. For example, in $H_2O^+$, the doubly occupied $1s$ orbital on the O-atom and one virtual orbital are removed, leaving FCI to only consider distributions of seven electrons over seven UNOs. For consistency, GHF calculations to build the GCM space were also performed in the same UNO basis, in addition to GCM calculations in the full basis. As a caveat we add that it must be ensured that UHF does not break point-group symmetry, because the UNO space otherwise will not lead to a $g$-tensor with the correct symmetry.

We further construct fictitious tetrahedral $CH_4^+$ and $CuF_4^{2-}$ species with orbitally degenerate ground states, whose SOC splitting through first order cannot be described by a single-determinant wave function. A CASSCF-type ansatz may thus appear to be required. We however show that a GCM space that takes orbital degeneracy into account can indeed handle the corresponding strong-correlation problem (at a mean-field cost) to yield qualitatively correct Kramers doublets and $g$-tensors.

## 3. Results and Discussion

Before presenting results for the test set, we exemplarily discuss various aspects of the GCM strategy for the $H_2O^+$ radical. This molecule (and all others) lies in the $xz$-plane, with $z$ being a rotational axis, see Figure 1.

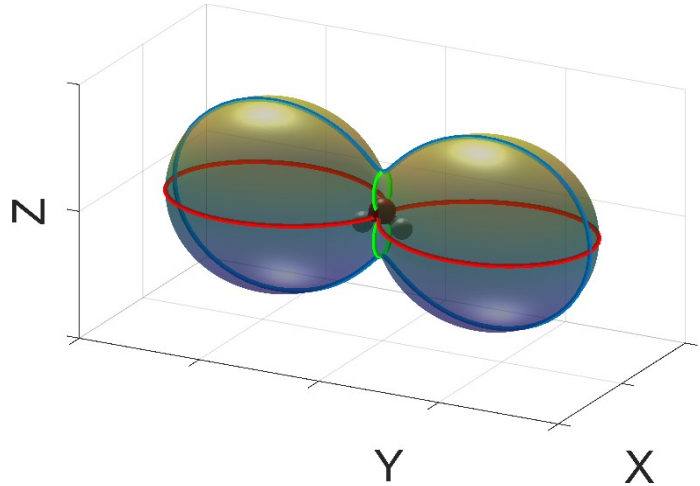

Figure 1: GHF energy surface, including SOC contributions, for $H_2O^+$. The distance from the origin (the center of nuclear charge) for an orientation **n** of the magnetic moment is proportional to the energy difference with respect to the global minimum $E_x = \langle \Phi_x | \hat{H} | \Phi_x \rangle$. See main text for further details; for energies, see Table 1 or Table 2. The energy curves for **n** lying in one of the coordinate planes ($xy$, red; $xz$, green; $yz$, blue) lie almost perfectly on the surface.

Taking the UHF solution $|\Xi_n\rangle$ ($\hat{H}_{SOC} = 0$) with spin quantised (projection $M = \tfrac{1}{2}$) along **n** = **x, y,** or **z** (all three UHF states are degenerate, due to spin symmetry) as the initial guess for GHF with the full Hamiltonian, $\hat{H} = \hat{H}_0 + \hat{H}_{SOC}$, yields $|\Phi_x\rangle$, $|\Phi_y\rangle$, and $|\Phi_z\rangle$, respectively. Even the unconstrained SCF algorithm straightforwardly achieves convergence onto these three GHF solutions.

The $C_{2v}^*$ double-group symmetry of $\hat{H}$ consists of combinations of the operations of the ordinary $C_{2v}$ group of $\hat{H}_0$ with corresponding spin rotations. The self-consistent symmetry (symmetries of the Fock operator, or, equivalently, the single-particle density matrix [43–45]) of each of the three GHF solutions is isomorphic to the magnetic group $C_{2v}^*(C_2^*)$ (this notation is explained in Ref. [46]). The nontrivial self-consistent symmetry operations listed in Table 1 are different between $|\Phi_x\rangle$, $|\Phi_y\rangle$, and $|\Phi_z\rangle$ (see Appendix B for further comments), but each of the three Kramers pairs $\{|\Phi_\mathbf{n}\rangle, |\bar{\Phi}_\mathbf{n}\rangle\}$ separately spans the two-dimensional irreducible representation (irrep) $E_{1/2}$ of the $C_{2v}^*$ group of the Hamiltonian ($E_{1/2}$ is the only fermionic irrep of $C_{2v}^*$ [47]). This indeed holds true for all $C_{2v}$ molecules in our test set (see footnote to Table 2 for a minor caveat on NO$_2$). Due to the described self-consistent symmetry, $\langle \hat{\mathbf{S}} \rangle$, $\langle \hat{\mathbf{L}} \rangle$ and $\langle \hat{\boldsymbol{\mu}} \rangle$ are all collinear (parallel or antiparallel), that is, $\langle \Phi_\mathbf{n} | \hat{\mathbf{S}} | \Phi_\mathbf{n} \rangle \propto \mathbf{n}$, $\langle \Phi_\mathbf{n} | \hat{\mathbf{L}} | \Phi_\mathbf{n} \rangle \propto \mathbf{n}$, and $\langle \Phi_\mathbf{n} | \hat{\boldsymbol{\mu}} | \Phi_\mathbf{n} \rangle \propto \mathbf{n}$. The same is true for the linear molecules in our set. For a general orientation **n**, these three vectors are not strictly collinear, because the respective constrained GHF solution has no self-consistent symmetry. However, due to the comparatively small orbital momentum $\langle \Phi_\mathbf{n} | \hat{\mathbf{L}} | \Phi_\mathbf{n} \rangle$, spin and magnetic moment are still very closely aligned.

Table 1: Energies of three GHF solutions in H$_2$O$^+$ (cc-pVTZ basis), nontrivial elements of the magnetic self-consistent symmetry group [isomorphic to $C_{2v}^*(C_2^*)$ in all cases] and *g*-shifts (in ppm) calculated based on Eq. (4) for all separate pairs $\{|\Phi_\mathbf{n}\rangle, |\bar{\Phi}_\mathbf{n}\rangle\}$. Values in bold type are identical to 3SCF results, see Appendix B.

|  | $|\Phi_x\rangle$ | $|\Phi_y\rangle$ | $|\Phi_z\rangle$ |
|---|---|---|---|
| $E$ | -75.658 031 344 | -75.658 031 054 | -75.658 031 289 |
| symmetries | $\exp(-i\pi\hat{S}_x) \times \hat{\sigma}_{yz}$ $\hat{\Theta} \times \exp(-i\pi\hat{S}_y) \times \hat{\sigma}_{xz}$ $\hat{\Theta} \times \exp(-i\pi\hat{S}_z) \times \hat{C}_2$ | $\exp(-i\pi\hat{S}_y) \times \hat{\sigma}_{xz}$ $\hat{\Theta} \times \exp(-i\pi\hat{S}_z) \times \hat{C}_2$ $\hat{\Theta} \times \exp(-i\pi\hat{S}_x) \times \hat{\sigma}_{yz}$ | $\exp(-i\pi\hat{S}_z) \times \hat{C}_2$ $\hat{\Theta} \times \exp(-i\pi\hat{S}_y) \times \hat{\sigma}_{xz}$ $\hat{\Theta} \times \exp(-i\pi\hat{S}_x) \times \hat{\sigma}_{yz}$ |
| $\Delta g_x$ | **-15** | -15475 | -15498 |
| $\Delta g_y$ | 1139 | **13751** | 1134 |
| $\Delta g_z$ | -11278 | -11248 | **3830** |

The GHF character $\gamma_\mathbf{n}$, Eq. (16), is rather small. For H$_2$O$^+$ we find $\gamma_x = 2.2 \times 10^{-5}$, $\gamma_y = 1.6 \times 10^{-5}$ and $\gamma_z = 2.2 \times 10^{-5}$. As $|\Phi_\mathbf{n}\rangle$ and $|\Xi_{-\mathbf{n}}\rangle$ belong to different components of $E_{1/2}$, they have zero overlap.

$$\gamma_n = 1 - |\langle \Phi_\mathbf{n} | \Xi_\mathbf{n} \rangle|^2 \tag{16}$$

It is not surprising that $|\Phi_x\rangle$, $|\Phi_y\rangle$ and $|\Phi_z\rangle$ are not degenerate, because there is no symmetry that relates any pair of these three states. The respective "magnetic anisotropy energy" (MAE) [19,48] is frequently computed to extract SOC contributions to zero-field splitting tensors of $S > \frac{1}{2}$ levels, usually in the frame of DFT [48–50]. The energy difference between GHF solutions in $H_2O^+$, $E_z - E_x \approx E_y - E_x \approx 0.2 \mu h$, is one or two orders of magnitude smaller than MAE in light $S = 1$ molecules [50], and it is about two orders of magnitude smaller than the energy dispersion associated with spin-rotations performed on self-consistent GHF solutions, without letting the spin-rotated state relax (see later section on diagonalisation in a spin-rotation manifold). The significance of MAE for $S = \frac{1}{2}$ g-tensors is not known [21], but shall be elucidated later in this section.

The dependence of the GHF energy, $E(\mathbf{n}) = \langle \Phi_\mathbf{n} | \hat{H} | \Phi_\mathbf{n} \rangle$, on the magnetic-moment orientation $\mathbf{n}$ (see Table 1 and Table 2) is very accurately described by an MAE-tensor $\mathbf{M}$, Eq. (17),

$$E(\mathbf{n}) = \mathbf{n}^T \cdot \mathbf{M} \cdot \mathbf{n} , \qquad (17)$$

with $\mathbf{M} = \mathrm{diag}(E_x, E_y, E_z)$. When parameterising $\mathbf{n}$ in terms of polar angles $(\vartheta, \phi)$,

$$\mathbf{n} = (\cos\phi \sin\vartheta, \sin\phi \sin\vartheta, \cos\vartheta)^T , \qquad (18)$$

the MAE function takes the explicit form of Eq. (19),

$$E(\vartheta, \phi) = (E_x \cos^2\phi + E_y \sin^2\phi) \sin^2\vartheta + E_z \cos^2\vartheta . \qquad (19)$$

The surface defined by the distance of $E(\vartheta,\phi) - \min(E_x, E_y, E_z)$ from the coordinate origin is plotted in Figure 1 for $H_2O^+$. Curves connecting 100 data points (not shown individually) for rotating $\mathbf{n}$ in the three coordinate planes lie almost perfectly on the surface. We checked that this also holds for arbitrary $\mathbf{n}$ (data not shown). Note that a state with arbitrary $\mathbf{n} \neq \mathbf{x}, \mathbf{y}, \mathbf{z}$ is not in general a valid stationary point of the HF energy functional and cannot therefore be converged unless constraints are imposed. The suggestion by Jayatilaka [20] to fix $\mathbf{n}$ by including the Zeeman term, Eq. (2), self-consistently and gradually reducing its magnitude towards convergence, is not a similarly well-defined and black-box procedure as our present constrained optimisation.

Table 2: UHF and GCM(6,UHF) energies (in Hartree, h), excluding SOC. For GHF solutions $|\Phi_x\rangle$, $|\Phi_y\rangle$ and $|\Phi_z\rangle$, energy differences, including SOC, are given with respect to UHF. For GCM(6) and GCM(48) energy differences are given with respect to GCM(6,UHF). All differences in $10^{-6}$ h.

|  | UHF | $|\Phi_x\rangle$ | $|\Phi_y\rangle$ | $|\Phi_z\rangle$ | GCM (6,UHF) | GCM(6) | GCM(48)[a] |
|---|---|---|---|---|---|---|---|
| $CO^+$ | -112.303 992 73 | -2.77 | -2.77 | -2.69 | -112.317 264 61 | -2.73 | -3.04 $\tilde{D}=20$ |
| CN | -92.234 197 34 | -0.83 | -0.83 | -0.79 | -92.250 994 18 | -0.85 | -2.07 $\tilde{D}=20$ |
| MgF | -299.142 482 33 | -15.53 | -15.53 | -15.54 | -299.142 572 94 | -15.55 | -15.55 $\tilde{D}=12$ |
| $NO_2$[b] | -204.108 535 36 | -7.47 | -7.54 | -7.59 | -204.114 929 77 | -8.09 | -8.12 $\tilde{D}=20$ |
| $NF_2$ | -253.260 924 96 | -15.30 | -15.18 | -15.22 | -253.265 534 18 | -15.33 | -15.33 $\tilde{D}=16$ |
| $CO_2^-$ | -187.651 597 05 | -5.26 | -5.25 | -5.27 | -187.654 803 84 | -5.27 | -5.27 $\tilde{D}=14$ |
| $O_3^-$ | -224.438 474 81 | -12.19 | -11.50 | -11.45 | -224.445 875 60 | -12.38 | -12.42 $\tilde{D}=20$ |
| $H_2O^+$ | -75.658 027 77 | -3.58 | -3.29 | -3.52 | -75.661 754 46 | -3.48 | -3.48 $\tilde{D}=12$ |
| $H_2CO^+$ | -113.566 339 13 | -3.46 | -3.40 | -3.18 | -113.573 102 53 | -3.34 | -3.40 $\tilde{D}=20$ |

[a] $\tilde{D}$ denotes the reduced basis dimension after removal of near linear degeneracies in GCM(48). [b]There is a slightly lower UHF solution in $NO_2$ (E = -204.10855905 h) that breaks $C_{2v}$ symmetry. If the latter UHF solution is used as an initial guess for GHF (including SOC), then each of the three GHF Kramers pairs breaks $C_{2v}^*$ symmetry, which would require including symmetry partners in the GCM basis to obtain a rhombic g-tensor. Our analysis is for simplicity based on symmetry-conserving UHF and GHF solutions.

A surface plot visually indistinguishable from Figure 1 would be obtained if **n** defined the orientation of $\langle \hat{\mathbf{S}} \rangle$ rather than $\langle \hat{\boldsymbol{\mu}} \rangle$, due to the small relative magnitude of $\langle \hat{\mathbf{L}} \rangle$. Yet the latter is quantitatively important for g-shifts (for several of the systems in our test set, separate spin- and orbital-momentum contributions were analysed in Ref. [9]). The MAE surface defines a non-degenerate GHF manifold, that is, a set of states that are nearly degenerate and unrelated by symmetry, except for opposite orientations **n** and (-**n**), which define a Kramers pair, $\{|\Phi_\mathbf{n}\rangle, |\Phi_{-\mathbf{n}}\rangle\}$.

Energy profiles for additional systems (CN, $CO_2^-$ and MgF) are displayed in Figure 2. As strongly suggested by these surfaces, fully unconstrained GHF stability analyses [51] on $|\Phi_x\rangle$, $|\Phi_y\rangle$ and $|\Phi_z\rangle$ show that only the lowest-energy solution is a minimum, most likely the global minimum, while the remaining ones are saddle points. For $H_2O^+$, $|\Phi_x\rangle$ is the global minimum, and $|\Phi_z\rangle$ and $|\Phi_y\rangle$ are identified as first- and second-order saddle points by the respective number of negative orbital-Hessian eigenvalues (with small magnitudes on the order of $10^{-8}$ a.u.). In addition, there are two small positive eigenvalues for $|\Phi_x\rangle$ and one for $|\Phi_z\rangle$. For $H_{SOC} \to 0$, Hessian eigenvectors belonging to small-magnitude eigenvalues become zero-modes (Goldstone modes) that signal broken continuous symmetry (spin-rotational symmetry about axes $x$ or $y$ for $|\Phi_z\rangle$, etc.). On the other hand, as noted, a constrained solution $|\Phi_n\rangle$ for an arbitrary **n** is not a stationary point. In the limit $\hat{H}_{SOC} \to 0$, the size of the MAE surface shrinks to zero and the GHF manifold becomes a spin-rotation manifold of the UHF solution that spans states $|S, M = \pm\tfrac{1}{2}\rangle$, with $S = \tfrac{1}{2}, \tfrac{3}{2}, ..., S_{max}$. Diagonalisation of $\hat{H}_0$ in this space yields pure spin states and thus amounts to PAV spin projection.

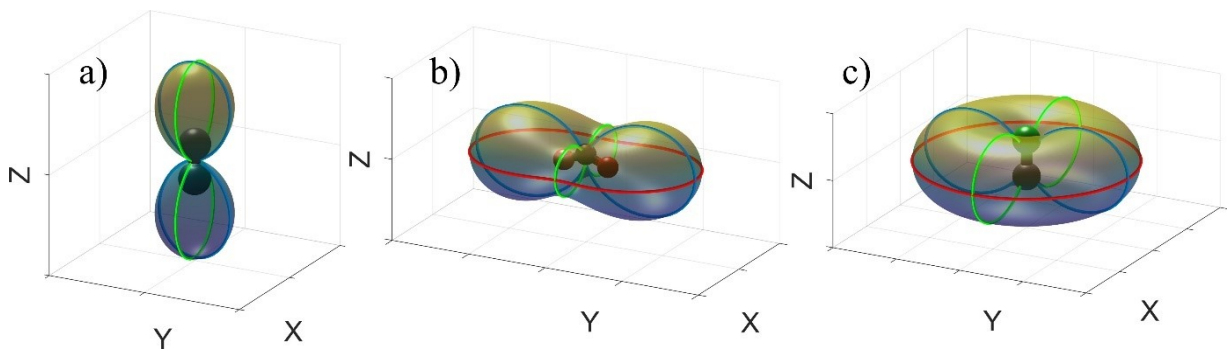

Figure 2: GHF energy surfaces (including SOC contributions) for CN (a), $CO_2^-$ (b) and MgF (c). For details, see caption to Figure 1; for energies, see Table 2.

Table 1 also includes *g*-tensors calculated from Eq. (4) for each of the three separate pairs $\{|\Phi_n\rangle, |\bar{\Phi}_n\rangle\}$. Due to the described symmetry-conserving properties of each Kramers pair, the three different respective *g*-tensors are correctly diagonal in the axis system of Figure 1. We explain in Appendix B why the principal component belonging to the magnetic-moment or spin direction coincides with the 3SCF result of $g_n = |\langle \Phi_n | \mathbf{n} \cdot \hat{\boldsymbol{\mu}} | \Phi_n \rangle|$. However, the two orthogonal components are entirely unreasonable when compared to 3SCF (or GCM, FCI, or experiment, see below). This problem was traced to spin contamination, quantified by the difference between $\langle \hat{\mathbf{S}}^2 \rangle$ and its ideal value of $S(S+1)$ in the absence of SOC [3]. Indeed, even when $\hat{H}_{SOC} = 0$, spin contamination yields an anisotropic *g*-tensor with non-zero *g*-shifts. As an

example, we obtain $\Delta g_z = 0$ and $\Delta g_x = \Delta g_y = -15433\text{ppm}$ for $H_2O^+$ from Eq. (4) based on the UHF solution $|\Xi_z\rangle$. This error is of the same magnitude as experimental *g*-shifts. Cherry et al. [3] proposed to subtract the quantity $\frac{g_e}{2}\left[\langle \mathbf{S}^2 \rangle - S(S+1)\right]$ (which must be multiplied by an appropriate phase factor) from cross-elements $\langle \Phi_\mathbf{n} | \mathbf{n}' \cdot \hat{\boldsymbol{\mu}} | \bar{\Phi}_\mathbf{n} \rangle$ in Eq. (4), where $\mathbf{n}' \cdot \mathbf{n} = 0$. By construction, this correction yields exactly $\mathbf{g} = g_e$ for $\hat{H}_{SOC} \to 0$, provided that $|\Phi_\mathbf{n}\rangle$ becomes a real UHF determinant $|\Xi_\mathbf{n}\rangle$ (with no orbital momentum) in this limit. Such an *a posteriori* correction would not be adequate for strong SOC, when spin is not an approximate symmetry anymore.

An arguably more elegant and further-reaching strategy is suggested by the realisation that $\mathbf{g} = g_e$ also results if contaminating spin states are eliminated from $|\Xi_\mathbf{n}\rangle$. One of the simplest approximate spin-projection schemes is a diagonalisation of $\hat{H}_0$ in the basis of six determinants oriented along octahedral directions, $\mathbf{n} = \pm \mathbf{x}, \pm \mathbf{y}, \pm \mathbf{z}$. For all molecules in the present set, this yields $\Delta g \ll 1\text{ppm}$ (in some cases, $\Delta g$ is as small as $10^{-6}$ ppm). In other words, an octahedral grid of UHF spin orientations, here called GCM(6,UHF), affords essentially pure $S = \frac{1}{2}$ doublets. Although this observation prompted us to proceed similarly based on three GHF Kramers doublets $\{|\Phi_\mathbf{n}\rangle, |\bar{\Phi}_\mathbf{n}\rangle\}$, $\mathbf{n} = \mathbf{x}, \mathbf{y}, \mathbf{z}$, with SOC included self-consistently, in what we call GCM(6), it is not entirely correct to motivate GCM(6) in terms of an approximate restoration of spin symmetry to remove spin-contamination errors in the *g*-tensor. We shall comment on this latter aspect in a later section on diagonalisation in a manifold of spin rotations.

Table 3 compares *g*-shifts from our 3SCF and GCM(6) calculations and the more dense sampling of GCM(48) (explained below) with 3SCF data from Ref. [20] and experimental values. Various sources may contribute to the (generally modest) differences between the two 3SCF sets, e.g., small differences in molecular structures, or the treatment of relativistic effects. While we employ only an effective one-electron SOC operator (Eq. (7)), a larger number of relativistic terms were explicitly considered in Ref. [20]. A trend favoring GCM(6) over 3SCF in terms of agreement with experimental data cannot be discerned, because predictions from both methods are quite similar.

Table 3: Experimental g-shifts (in ppm; molecules lie in the yz-plane and z is a rotational symmetry axis) are compared to 3SCF results from Ref. [20] and to our present 3SCF, GCM(6) and GCM(48) results.

|  | $\Delta g$ | Exp. | 3SCF [20] | 3SCF | GCM(6) | GCM(48) |
|---|---|---|---|---|---|---|
| $CO^+$ | ⊥ | -2400/-3000 | -2798 | -2992 | -2260 | -2260 |
|  | ∥ | -1200/-1400 | -42 | 0 | -2 | -1 |
| CN | ⊥ | -2000 | -1983 | -2003 | -1738 | -1736 |
|  | ∥ | -700 | -63 | -1 | -6 | -1 |
| MgF | ⊥ | -1300 | -1314 | -1839 | -1814 | -1815 |
|  | ∥ | -300 | 20 | -2 | -2 | -2 |
| $NO_2$ | x | 3900 | 3368 | 4762 | 5239 | 5240 |
|  | y | -11300 | -11008 | -12642 | -12838 | -12839 |
|  | z | -300 | -623 | -1070 | -281 | -280 |
| $NF_2$ | x | -100 | -447 | -747 | -258 | -258 |
|  | y | 6200 | 5649 | 6112 | 6556 | 6558 |
|  | z | 2800 | 2892 | 3110 | 3412 | 3410 |
| $CO_2^-$ | x | 700/800 | 805 | 1190 | 1396 | 1397 |
|  | y | -4800/-5070 | -4767 | -5976 | -5987 | -5988 |
|  | z | -500/-710 | -571 | -741 | -450 | -450 |
| $O_3^-$ | x | 200/1300 | -597 | -743 | -307 | -307 |
|  | y | 14700–16200 | 17933 | 21108 | 23530 | 23529 |
|  | z | 10000/9700 | 10945 | 13650 | 15755 | 15755 |
| $H_2O^+$ | x | 200 | -229 | -15 | -41 | -41 |
|  | y | 18800 | 12704 | 13751 | 15706 | 15718 |
|  | z | 4800 | 3306 | 3830 | 4078 | 4078 |
| $H_2CO^+$ | x | 4600 | 5472 | 5871 | 6020 | 6016 |
|  | y | -800 | 927 | 799 | 233 | 234 |
|  | z | 200 | 2976 | 2977 | 4629 | 4628 |

To probe if GCM(6) sufficiently samples the manifold of orientations of $\langle\hat{\boldsymbol{\mu}}\rangle$, we also generated a larger basis by constraining $\langle\hat{\boldsymbol{\mu}}\rangle$ to point towards the vertices of an icosahedron. This icosahedral grid was arbitrarily rotated with respect to the molecule to avoid any equivalence of grid points due to molecular symmetry. However, only six (instead of twelve) constrained GHF calculations are needed, because pairs of GHF solutions with $\langle\hat{\boldsymbol{\mu}}\rangle$ pointing to diametrically opposite vertices of the icosahedron are Kramers pairs and thus generated by $\hat{\Theta}$. This basis of dimension 12 does not span a representation of the molecular double group ($C_{2v}^*$ or $C_{\infty v}^*$). Therefore, a Kramers doublet from diagonalisation in this space does not span an irrep, and the g-tensor will deviate (albeit slightly) from its correct symmetry. For $C_{2v}$ molecules, we therefore include all determinants ("symmetry partners") obtained from the application of one of the operations $\{\hat{E}, \exp(-i\pi\hat{S}_z)\times\hat{C}_2, \exp(-i\pi\hat{S}_x)\times\hat{\sigma}_{yz}, \exp(-i\pi\hat{S}_y)\times\hat{\sigma}_{xz}\}$ of $C_{2v}^*$ ($\hat{E}$ is the identity; the omitted remaining four operations of $C_{2v}^*$ are associated with an additional $2\pi$ spin rotation about an arbitrary axis, which just introduces a factor of (–1)). This amounts to PAV restoration of $C_{2v}^*$ symmetry. (Note that PAV restoration of point-group symmetry in the frame of NOCI was extensively discussed in a different context in Ref. [52].) In linear molecules, we are only interested in an approximate restoration of $\hat{J}_z = \hat{S}_z + \hat{L}_z$ symmetry. To this end, $\hat{U} = \exp(-i\phi \hat{J}_z)$ rotates the twelve original determinants by angles $\phi = 2\pi k/4$, $k = 0,1,2,3$, about the molecular axis. Although this corresponds only to explicit $\hat{C}_4^*$ restoration, it still guarantees perfect axiality, $\mathbf{g} = \mathrm{diag}(g_\perp, g_\perp, g_\parallel)$. A lower threshold of $10^{-8}$ for the eigenvalues of the overlap matrix is applied to remove (near) linear dependencies in the GCM(48) basis. The reduced basis dimensions $\tilde{D}$ are collected in Table 2. For $H_2O^+$, all $g$-tensor components from GCM(48) differ by only a few ppm from GCM(6). Even with denser spherical grids (we tested a Lebedev-Laikov grid [53] with 146 points) and a smaller threshold of $10^{-10}$ to define linear dependencies, there are still only $\tilde{D} = 12$ linearly-independent states and the $g$-tensor remains unchanged to within ppm accuracy. Overall, at least with regard to the $g$-tensor, the nearly degenerate manifold of spin or magnetic-moment orientations is sufficiently sampled by GCM(6). This is even true in cases with significant spin contamination. For example, the CN radical has $\langle \hat{\mathbf{S}}^2 \rangle = 0.9949$ in UHF (the ideal value is $S(S+1) = \frac{3}{4}$), versus $\langle \hat{\mathbf{S}}^2 \rangle = 0.7577$ for $H_2O^+$. For CN, the GCM(48) energy is lower by $> 10^{-6}$ h compared to GCM(6), while the energy difference is $< 10^{-9}$ h for $H_2O^+$, but the GCM(6) $g$-shifts are similarly well converged in both cases.

As to the relevance of energy differences between $|\Phi_x\rangle, |\Phi_y\rangle, |\Phi_z\rangle$, we checked that they have a negligible effect on $g$-tensors by setting the diagonal elements of the GCM(6) matrix to zero, which causes changes of $\ll 1\mathrm{ppm}$. Even artificially scaling up MAE in GCM(6) by two orders of magnitude has generally only a very small effect. One may even entirely exclude $\hat{H}_{\mathrm{SOC}}$ from the NOCI problem without affecting $g$-tensors to any relevant degree, as long as GHF solutions are adapted to SOC. In other words, $\hat{H}_{\mathrm{SOC}}$ must be included self-consistently in GHF calculations but may be dropped in the subsequent NOCI step. We however cannot exclude that in systems with strong SOC, a consideration of MAE and the related inclusion of $\hat{H}_{SOC}$ in the NOCI step will make a sizable difference. We therefore like to emphasise that MAE does not pose any conceptual problem in the GCM approach.

**Diagonalisation in a manifold of spin rotations**. The fact that spin contamination strongly affects *g*-tensors calculated from a single GHF Kramers doublet $\{|\Phi_n\rangle, |\bar{\Phi}_n\rangle\}$ [3] suggests to proceed like in a projective elimination of contaminating spin states. With a projector parameterised by three Euler angles [54,55], spin projection corresponds to an *SU*(2) integration. However, as SOC breaks spin symmetry, direct projection to obtain a pure $S = \frac{1}{2}$ state from $|\Phi_n\rangle$ is not adequate. It is still possible to apply the spin rotations that make up the grid-discretised projector (where each rotation operation is weighted by an element of the Wigner rotation matrix, which is not needed here) to span a spin-rotation basis from $|\Phi_n\rangle$. (Incidentally, this approach is similar in spirit to applying separate localised spin rotations to molecular fragments to span a GCM basis that correctly describes dissociation, where local spin is not a good quantum number, except at an infinite separation of fragments [56].) In the limit $\hat{H}_{SOC} \to 0$, diagonalisation in this basis will yield a pure $S = \frac{1}{2}$ state that could be obtained directly by applying the respective spin projector to the HF wave function; higher-lying NOCI solutions represent contaminating spin states.

The following efficient *SU*(2) integration grids appear useful: a Lebedev-Laikov [53]/trapezoid grid combination [57], grids based on the proper symmetry operations of symmetric polyhedra [58], or Coxeter grids [59]. Figure 3 illustrates the second option [57]. A selected GHF solution in $H_2O^+$ is arbitrarily spin-rotated, so that the resulting GHF-type determinant does not have any self-consistent symmetry. This excludes orthogonality with respect to any of the other determinants in the basis, except the respective Kramers partner, which facilitates the calculation of matrix elements. Next, spin rotations corresponding to the 60 proper symmetry operations of an arbitrarily oriented icosahedron are applied to this reference determinant. Lastly, Kramers partners are included to restore time-reversal symmetry. The resulting basis of dimension $D = 120$ contains many near linear dependencies (that we remove) and does not rigorously span a $C_{2v}^*$ representation. We however deliberately choose to not include point-group partners to assess the diagonalisation of $\hat{H}_0 + \hat{H}_{SOC}$ in a manifold of spin rotations independent of molecular symmetry.

Table 4: g-shifts (in ppm) from a diagonalisation of $\hat{H}_0 + \hat{H}_{SOC}$ in a manifold of spin rotations of GHF solutions $|\Phi_x\rangle$, $|\Phi_y\rangle$ or $|\Phi_z\rangle$. $\tilde{D}$ denotes the reduced basis dimension (see main text).

|   |   | GCM(**x**,120) | GCM(**y**,120) | GCM(**z**,120) |
|---|---|---|---|---|
| $CO^+$ | $x$ | -2001 | -1689 | -1689 |
|  | $y$ | -1689 | -2001 | -1689 |
|  | $z$ | -1 | -1 | -1 |
|  | $\tilde{D}$ | 70 | 70 | 76 |
| CN | $x$ | -1336 | -1326 | -1337 |
|  | $y$ | -1326 | -1336 | -1337 |
|  | $z$ | -1 | -1 | -1 |
|  | $\tilde{D}$ | 70 | 70 | 70 |
| MgF | $x$ | -1701 | -1678 | -1678 |
|  | $y$ | -1678 | -1701 | -1678 |
|  | $z$ | -2 | -2 | -2 |
|  | $\tilde{D}$ | 38 | 38 | 38 |
| $NO_2$ | $x$ | 4043 | 3358 | 3335 |
|  | $y$ | -10625 | -10438 | -10627 |
|  | $z$ | 36 | 34 | -670 |
|  | $\tilde{D}$ | 64 | 70 | 70 |
| $NF_2$ | $x$ | -495 | -89 | -92 |
|  | $y$ | 5837 | 5391 | 5838 |
|  | $z$ | 3004 | 3009 | 2816 |
|  | $\tilde{D}$ | 64 | 64 | 64 |
| $CO_2^-$ | $x$ | 1116 | 1305 | 1305 |
|  | $y$ | -5118 | -5161 | -5118 |
|  | $z$ | -290 | -290 | -545 |
|  | $\tilde{D}$ | 64 | 64 | 64 |
| $O_3^-$ | $x$ | -495 | -99 | -115 |
|  | $y$ | 16130 | 14438 | 16146 |
|  | $z$ | 9352 | 9380 | 8820 |
|  | $\tilde{D}$ | 76 | 76 | 76 |
| $H_2O^+$ | $x$ | -27 | -35 | -53 |
|  | $y$ | 15831 | 13460 | 15836 |
|  | $z$ | 3984 | 4007 | 3608 |
|  | $\tilde{D}$ | 44 | 38 | 38 |
| $H_2CO^+$ | $x$ | 5466 | 5619 | 5632 |
|  | $y$ | 1 | 574 | 7 |
|  | $z$ | 4743 | 4740 | 2668 |
|  | $\tilde{D}$ | 64 | 64 | 64 |

The g-shifts resulting from GCM(**x**,120), GCM(**y**,120) and GCM(**z**,120), which work on a basis of GHF solutions $|\Phi_x\rangle$, $|\Phi_y\rangle$ and $|\Phi_z\rangle$, respectively, are given in Table 4. The

deviations of *g*-tensors from perfect rhombicity are in all cases extremely small (not detailed) and the dependence on the choice of the GHF reference is modest. The results are quite similar to GCM(6) and show that a Kramers doublet that yields a reasonable *g*-tensor can be obtained based on just a single GHF solution. Somewhat unfortunately, GCM(**x**,120) and GCM(**y**,120) noticeably break axial symmetry in linear molecules (see, e.g., CO$^+$). This problem is absent in GCM(**z**,120). All results are well converged with respect to the grid size (larger Lebedev-Laikov/trapezoid grids cause changes <1 ppm), grid orientation, as well as the arbitrary spin rotation of the GHF solution (cf. Figure 3b). In most cases, even a smaller grid, e.g., replacing the icosahedron in Figure 3c by an octahedron, is sufficient. We however found that no reasonable results can be obtained when restricting the spin-rotation basis to just six states.

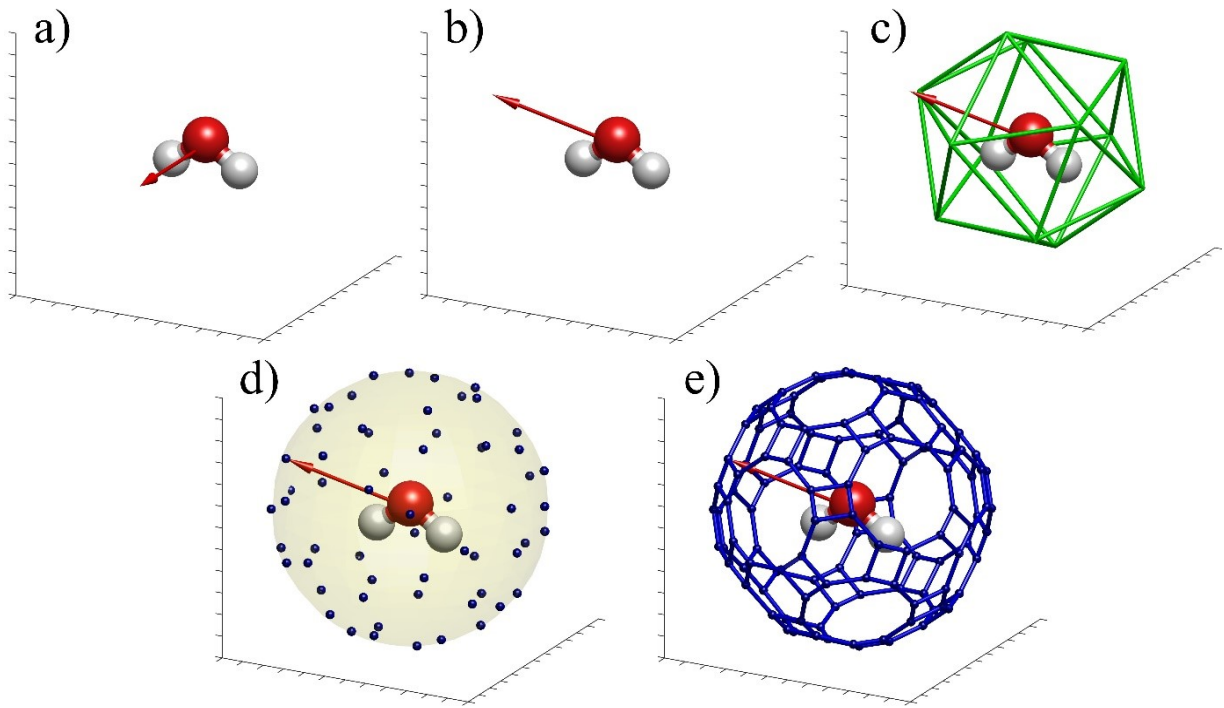

Figure 3: Spanning a spin-rotation NOCI basis. A GHF solution with spin in some direction (the red arrow in (a) represents $\langle \Phi_x | \hat{\mathbf{S}} | \Phi_x \rangle \propto \mathbf{x}$ ) is arbitrarily spin rotated (b). An arbitrarily oriented icosahedron (c) defines a spin-projection grid in terms of spin rotations corresponding to proper symmetries of the icosahedron. This yields 60 states with spin orientations marked by blue points on a sphere (d). Including Kramers partners doubles the number of determinants; points in (e) are connected by lines to guide the eye.

The energy dispersion on the diagonal of the NOCI matrix **H** is about two orders of magnitude larger than energy differences for self-consistent (fully relaxed) GHF states employed in GCM(6) or GCM(48), and crucially important. Neglecting the energy dispersion affords an almost pure $S = \tfrac{1}{2}$ NOCI solution (with $\Delta g < 10^{-3}$ ppm ), which would strictly become a pure spin state if we also neglected the unimportant off-diagonal elements of $\hat{H}_{\text{SOC}}$. GCM based on several optimised GHF solutions and GCM in a spin-rotation space of a single GHF solution can overall be regarded as two complementary approaches that yield rather similar *g*-tensors for our test set.

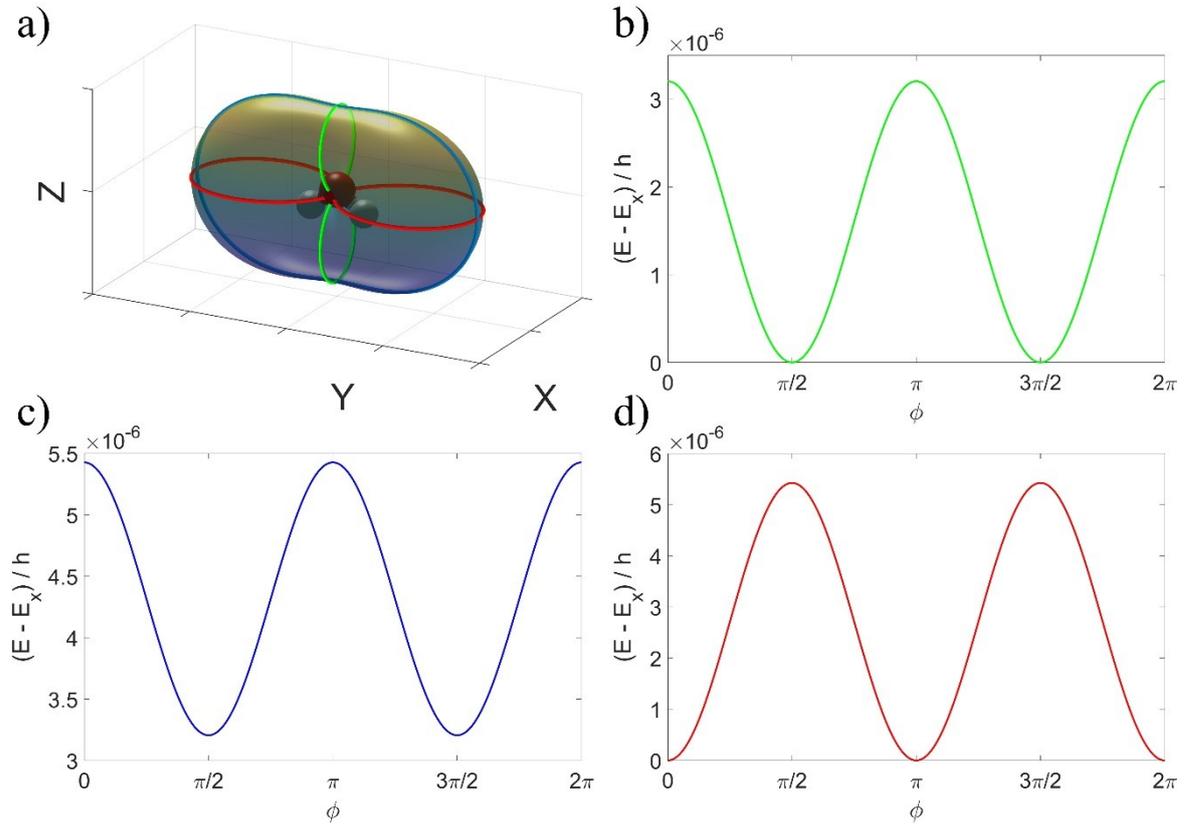

Figure 4: The MAE surface (cf. caption to Figure 1) for spin rotations of the GHF solution $|\Phi_x\rangle$ in $H_2O^+$ is plotted in (a). Due to the transformation properties of spin operators under spin rotations, the direction dependence of the expectation value of $\hat{H}_{\text{SOC}}$ (Eq. (7)) is exactly described by Eq. (19), where $E_x \equiv \langle \Phi_x | \hat{H} | \Phi_x \rangle = -75.65802777$ (Table 2), $E_y \equiv \langle \Phi_x | e^{+i\frac{\pi}{2}\hat{S}_z} \hat{H} e^{-i\frac{\pi}{2}\hat{S}_z} | \Phi_x \rangle = -75.65802592$, and $E_z \equiv \langle \Phi_x | e^{+i\frac{\pi}{2}\hat{S}_y} \hat{H} e^{-i\frac{\pi}{2}\hat{S}_y} | \Phi_x \rangle = -75.65802814$. Curves (b), (c), and (d) are shown in the same colors as in (a) and have a perfect cosine form [in (b), $\phi$ is the angle with respect to the *z*-axis for a spin rotation about the *y*-axis, yielding $|\Phi_x\rangle$ for $\phi = \frac{\pi}{2}$ and $\phi = \frac{3}{2}$, etc.].

**Comparison with FCI.** It is difficult to judge the suitability of experimental *g*-values for a direct comparison with quantum-chemical calculations that cannot consider all (possibly minor) complicating aspects. While magnetic contributions from molecular rotations must be subtracted from gas phase data to yield intrinsic electronic properties, EPR measurements in the solid state are influenced by the environment (commonly a noble-gas matrix), which is another source of uncertainties. Besides, molecular vibrations may affect *g*-tensors, but are rarely considered (the present work is no exception). As an unambiguous benchmark for the intrinsic accuracy of 3SCF or GCM, we therefore ran FCI calculations on CN, $CO^+$, $H_2O^+$. The number of electronic configurations was reduced by working with an UNO active space, with frozen 1*s* cores, see Theory section. The effects of the basis truncation (UNO or FC-UNO) in 3SCF and GCM(6) are detailed in Table 5. With few exceptions, most notably the dramatic change in the $g_y$ component in $NO_2$, UNO (or FC-UNO) and full orbital-space results are reasonably similar.

Table 5: 3SCF and GCM(6) *g*-shifts with either the full cc-pVTZ basis, an UNO active space, or an UNO active space with a frozen 1*s* cores (FC-UNO).

|  |  | Full | | UNO | | FC-UNO | |
|---|---|---|---|---|---|---|---|
|  | $\Delta g$ | 3SCF | GCM(6) | 3SCF | GCM(6) | 3SCF | GCM(6) |
| $CO^+$ | $\perp$ | -2992 | -2260 | -2777 | -1972 | -2671 | -1902 |
|  | $\parallel$ | 0 | -2 | 0 | -2 | 0 | -1 |
| CN | $\perp$ | -2003 | -1738 | -1866 | -1349 | -1826 | -1313 |
|  | $\parallel$ | -1 | -6 | 0 | -5 | 0 | -5 |
| $NO_2$ | *x* | 4762 | 5239 | 4493 | 4447 | 4349 | 4322 |
|  | *y* | -12642 | -12838 | -3566 | -3738 | -3569 | -3742 |
|  | *z* | -1070 | -281 | -356 | -270 | -346 | -266 |
| $NF_2$ | *x* | -747 | -258 | -571 | -192 | -440 | -150 |
|  | *y* | 6121 | 6556 | 6203 | 6379 | 4975 | 5144 |
|  | *z* | 3110 | 3412 | 3023 | 3170 | 2961 | 3088 |
| $CO_2^-$ | *x* | 1190 | 1396 | 1457 | 1440 | 1407 | 1406 |
|  | *y* | -5976 | -5987 | -2223 | -2371 | -2222 | -2370 |
|  | *z* | -741 | -450 | -314 | -221 | -317 | -224 |
| $O_3^-$ | *x* | -743 | -307 | -392 | -175 | -355 | -161 |
|  | *y* | 21108 | 23530 | 19567 | 20234 | 19177 | 19828 |
|  | *z* | 13650 | 15755 | 12074 | 12384 | 11750 | 12002 |
| $H_2O^+$ | *x* | -15 | -41 | -13 | -35 | -9 | -31 |
|  | *y* | 13751 | 15706 | 13584 | 14256 | 13104 | 13732 |
|  | *z* | 3830 | 4078 | 3958 | 4161 | 3964 | 4167 |
| $H_2CO^+$ | *x* | 5871 | 6020 | 5654 | 5741 | 5652 | 5743 |
|  | *y* | 799 | 233 | 697 | 224 | 696 | 223 |
|  | *z* | 2977 | 4629 | 3427 | 4519 | 3432 | 4524 |

For the selected three molecules, we compare FC-UNO 3SCF and GCM(6) to FCI in Table 6. A clear trend in favor of either 3SCF or GCM(6) is not apparent, but the semi-quantitative agreement with FCI is encouraging and corroborates that both approaches are reasonable. GCM and FCI also agree reasonably well in the energetic stabilisation of the ground state brought about by SOC. Specifically, the difference between GCM(6,UHF) and GCM(6) energies is $1.26 \times 10^{-6}$ h for $CO^+$, $3.05 \times 10^{-7}$ h for CN and $2.04 \times 10^{-6}$ h for $H_2O^+$, versus differences in the "nonrelativistic" and "relativistic" FCI energies of $1.03\times10^{-6}$ h, $2.55\times10^{-7}$ h and $2.01\times10^{-6}$ h, respectively. In summary, GCM(6) predicts qualitatively correct $g$-tensors at a mean-field cost. It should be noted however that the lack of dynamical electron correlation in GCM(6) would represent a limitation for the description of other systems not treated here, where GCM(6) may not be expected to provide satisfactory results.

Table 6: Experimental $g$-shifts (in ppm) are compared to GCM(6) and FCI in an FC-UNO active space. The last two columns contain the FCI energies with or without SOC.

|  | $\Delta g$ | Exp. | 3SCF | GCM(6) | FCI |
|---|---|---|---|---|---|
| $CO^+$ | $\perp$ | -2400/-3000 | -2671 | -1902 | -2022 |
|  | $\parallel$ | -1200/-1400 | 0 | -1 | -1 |
| CN | $\perp$ | -2000 | -1826 | -1313 | -1610 |
|  | $\parallel$ | -700 | 0 | -5 | -1 |
| $H_2O^+$ | $x$ | 200 | -9 | -31 | 4 |
|  | $y$ | 18800 | 13104 | 13732 | 12819 |
|  | $z$ | 4800 | 3964 | 4167 | 3978 |

**Orbitally degenerate states.** We lastly demonstrate how a GCM approach can be applied to nonrelativistic ground states with orbital degeneracy, where a first-order spin-orbit splitting leads to $g$-shifts that are several orders of magnitude larger than in the foregoing examples. As two rather artificial yet illustrative examples with true orbital degeneracy, we construct tetrahedral $CH_4^+$ (bond length 1.07 Å) and $CuF_4^{2-}$ species (1.92 Å) which would both be subject to Jahn-Teller distortions. Chibotaru und Ungur studied $T_d \rightarrow D_{4h}$ and $T_d \rightarrow D_{2h}$ deformations in $CuCl_4^{2-}$ [23]. Using a CASSCF-SOSI procedure, they found that $g_x g_z / g_y$ (Eq. (5)) is negative at (and near) a $T_d$ structure. We only indicate how near orbital degeneracies for small structural deviations from perfect $T_d$ symmetry could be treated in the frame of GCM.

Tetrahedral $CH_4^+$ has a nonrelativistic $^2T_1$ ground state, as expected from the $t_1$ highest occupied molecular orbital (HOMO) in neutral $CH_4$. Orbital degeneracy constitutes a multi-reference problem. UHF indeed strongly breaks symmetry, $T_d \rightarrow C_{3v}$, where one H atom has Mulliken charge and spin densities of +0.46 and +0.44, respectively, with +0.27 and +0.02 on the remaining three H atoms. These calculations used an STO-6G basis to allow a direct comparison between NOCI and FCI. UNOs from a larger basis set are not adequate, because UHF breaks symmetry, and FCI solutions in a symmetry-broken UNO space would necessarily break symmetry too. SOC splits the six-fold degenerate $^2T_1$ level into a four-fold degenerate ground state (irrep $F_{1/2}$ in $T_d^*$ [47]) and a doubly-degenerate excited state. We are interested in the latter, which can be treated in terms of $\tilde{S} = \frac{1}{2}$, whereas the ground state formally represents an $\tilde{S} = \frac{3}{2}$ system. The described splitting pattern is in accord with the SOC splitting of an atomic term, $^2L \to L_{3/2} \oplus L_{1/2}$: the more-than half-filled valence shell of the central C atom in $CH_4^+$ leads to a negative effective SOC constant [60] that energetically favors $\tilde{S} = \frac{3}{2}$ (which could also be termed $\tilde{J} = \frac{3}{2}$) resulting from the coupling between $S = \frac{1}{2}$ and a pseudo-orbital momentum [2] $\tilde{L} = 1$. FCI predicts a splitting of $0.12 \times 10^{-3}$ h between $\tilde{S} = \frac{3}{2}$ and $\tilde{S} = \frac{1}{2}$ levels and yields an isotropic g-tensor for $\tilde{S} = \frac{1}{2}$, $g_x = g_y = g_z = 0.438$. From Eq. (5) we checked that the g-value is positive. Overall, this result suggests a qualitative explanation by the Landé formula,

$$g_J = \frac{3}{2} + \frac{S(S+1) - L(L+1)}{2J(J+1)} \ . \tag{20}$$

Setting $S = \frac{1}{2}$, and replacing $L$ by $\tilde{L} = 1$ and $J$ by $\tilde{S} = \frac{1}{2}$ in Eq. (20), yields $g_{\tilde{S}} = \frac{2}{3}$, in semi-quantitative agreement with FCI.

It is straightforward to obtain a result similar to FCI by diagonalising $\hat{H} = \hat{H}_0 + \hat{H}_{SOC}$ in a single-determinant basis that approximately spans the nonrelativistic $^2T_1$ ground state. This can be accomplished by generating all $T_d^*$ and Kramers partners of a symmetry-broken UHF solution with arbitrary spin orientation. The resulting perfectly isotropic g-tensor does not deviate by more than $\sim 0.05$ from the FCI result, slightly depending on the spin orientation in the respective UHF reference.

However, the described $T_d^*$ projection is ill-defined for a distorted structure. A simple solution to this problem reverts to the GCM approach of probing a manifold of $\langle \hat{\mathbf{\mu}} \rangle$ or $\langle \hat{\mathbf{S}} \rangle$ orientations, as proposed in previous sections. However, due to the multi-reference character, we must sample such manifolds separately for each of the four symmetry-broken solutions. Each such manifold could be followed adiabatically along a distortion coordinate. (On the other hand, working with a single manifold associated with the energetically favored UHF solution should be appropriate for strongly distorted structures.)

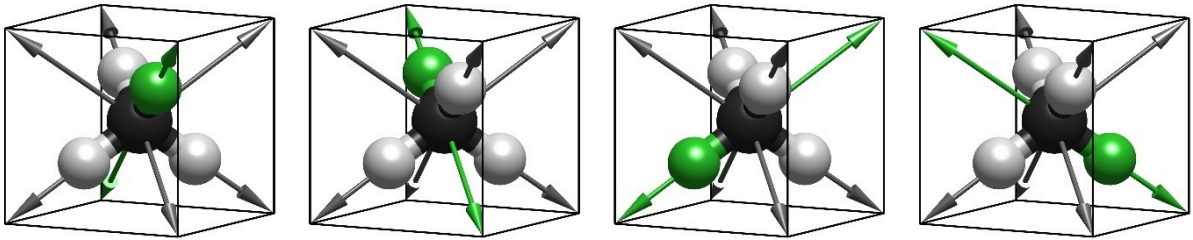

Figure 5: GCM(32) basis in tetrahedral $CH_4^+$. For each of the four equivalent symmetry-broken UHF solutions ($T_d \to C_{3v}$, the respective distinct H-atom is shown in green) a spherical manifold of spin orientations is sampled with eight grid points (vertices of a cube). Without SOC, all 32 states are degenerate, due to spin- and point-group symmetry. In the symmetry group of the full Hamiltonian (including SOC), the eight determinants represented by green arrows (spin parallel or antiparallel to the distinct C–H bond direction) and the remaining 24 determinants are separately related by operations of the double group or by time-reversal $\hat{\Theta}$. The basis of 32 states overall furnishes a representation of $T_d^*$ and maintains time-reversal symmetry.

For $CH_4^+$, we build a basis by orienting spin in each of the four equivalent symmetry-broken UHF solutions along one of the four C–H bond directions. Including Kramers partners, spin points to one of the eight vertices of a cube, see Figure 5. Diagonalisation in the full basis of dimension $D = 32$, which spans a representation of $T_d^*$, yields a splitting of $0.13 \times 10^{-3}$ h between $\tilde{S} = \frac{3}{2}$ and $\tilde{S} = \frac{1}{2}$ levels, and an isotropic $g = 0.471$, in reasonable agreement with FCI. Basing the described strategy on GHF (including SOC) instead of UHF (without SOC) does not cause significant changes.

The 3SCF approach is not suitable for the present problem, because it is intrinsically incapable of describing the coupling between spin and pseudo-orbital momentum. The same applies to working in a spin-rotation manifold of a single symmetry-broken UHF solution: GCM(8), in contrast to GCM(32), neither describes orbital degeneracy nor does it afford an isotropic $g$-tensor.

For $CuF_4^{2-}$, we used basis sets def2-SVP [61,62] on Cu and jun-cc-pVDZ [63] on F. The orbitally degenerate nature of $CuF_4^{2-}$ is signaled by UHF symmetry-breaking, $T_d \to C_{2v}$, where six degenerate UHF solutions differ in the distribution of charge- and spin-densities over two pairs of F-atoms. However, symmetry breaking is far less pronounced than in $CH_4^+$, energetic barriers separating the six configurations in $CuF_4^{2-}$ are of the same order of magnitude as SOC energies and consequently an arbitrary constrained GHF solution (including SOC) cannot generally be attributed to one of six manifolds. In other words, manifolds are not strictly separate. We can nevertheless pursue a very similar strategy as in $CH_4^+$. Successively taking one of the six UHF solutions with spin oriented along one of the four bond directions as initial guess, we constrain $\langle\hat{\mathbf{S}}\rangle$ along the given direction in GHF, which overall affords two sets of 12 GHF solutions related by $T^*$ (these are not two sets of 24 solutions related by $T_d^*$, because for the chosen constraint directions, GHF preserves a mirror symmetry, leaving two F atoms equivalent). Including Kramers partners, the respective GCM(48) yields an isotropic $g = -1.898$. The negative sign was established from Eq. (5). Chibotaru and Ungur calculated $g \approx -2$ for a tetrahedral configuration of $CuCl_4^{2-}$ [23], which provides confidence that our present result for $CuF_4^{2-}$ is at least semi-quantitatively correct. (Due to limitations of system size in our in-house code, we did not pursue calculations on $CuCl_4^{2-}$.) A space spanned by $T_d^*$ and Kramers partners of a single arbitrary GHF solution yields a very similar $g$-value, with only a slight dependence on the constraint orientation $\langle\hat{\mathbf{S}}\rangle$, but, as mentioned for $CH_4^+$, such a symmetry-projection strategy would not work for distorted structures. We add as a general caveat that in specific cases, the action of a double-group operation on a determinant can be equivalent to the action of $\hat{\Theta}$, so that duplicates must be removed from the basis spanned by double-group and Kramers partners. This complication can be avoided by choosing an orientation $\langle\hat{\mathbf{S}}\rangle$ unrelated to symmetry elements of the molecule. We lastly note that negative $g$-values are an indication of strong coupling between spin and orbital degrees of freedom, so that pseudospin is qualitatively different from true electronic spin [23]. A single determinant clearly cannot capture these features. Thus, just as in $CH_4^+$, 3SCF is inadequate to calculate $g$-tensors in a tetrahedral configuration of $CuF_4^{2-}$. Besides, 3SCF cannot in principle predict the sign of $g$-values. In contrast, our GCM approach demonstrates that such multi-reference problems can be handled qualitatively correctly at a mean-field cost.

## 4. Conclusions

We have explored different strategies for the calculation of molecular $g$-tensors based on non-orthogonal configuration interaction (NOCI) between GHF determinants. Two complementary strategies inspired by the generator coordinate method (GCM) were explored to span the basis: i) sampling a spherical manifold of $\langle\hat{\boldsymbol{\mu}}\rangle$ or $\langle\hat{\mathbf{S}}\rangle$ orientations in terms of nearly degenerate GHF solutions, including SOC self-consistently, ii) applying spin rotations to a single GHF solution.

For a set of small main-group radicals, the first approach requires just three GHF Kramers pairs. Predictions of $g$-tensors from the respective GCM(6) are similar to the common 3SCF strategy, but GCM(6) is conceptually more appealing, because it yields a single qualitatively correct Kramers pair to calculate the complete *g*-tensor, a feature that could become important when optimising geometries or calculating vibrational frequencies in the presence of SOC. A non-degenerate manifold can be densely sampled by constraining the directions of $\langle \hat{\boldsymbol{\mu}} \rangle$ or $\langle \hat{\mathbf{S}} \rangle$. The GCM approach is thus not restricted to three inequivalent GHF solutions that represent stationary points in rhombic molecules. This essentially removes the need to know the principal axes *a priori* and would straightforwardly allow to treat molecules with lower symmetry.

The non-degeneracy of the different GHF solutions is implicitly incorporated through NOCI, but we observed that the energy dispersion of the constrained manifold is essentially irrelevant for *g*-tensors, at least in our test set with weak SOC. On the other hand, when generating the basis from spin rotations of a single GHF solution, without relaxation, the energy dispersion is crucial to obtain reasonable *g*-tensors.

Working with an active space of UNOs permitted a rigorous evaluation of 3SCF and GCM against FCI. GCM agrees rather well with basis-set exact results from FCI (in an UNO space) but does not represent a quantitative improvement over 3SCF. On the other hand, two tetrahedral model systems have served to demonstrate that GCM is also applicable to orbitally degenerate systems with first-order SOC-induced level splitting and large *g*-shifts. While the 3SCF single-determinant approach intrinsically fails in these cases, GCM predicts qualitatively correct isotropic *g*-tensors, and even negative *g*-values in a $CuF_4^{2-}$ complex.

We believe that studies of larger systems with more pronounced SOC effects, particularly transition-metal complexes, would be worthwhile, although such systems pose additional challenges and may require capturing dynamic correlation effects, which the present GHF-based GCM approach is missing. Finally, NOCI in a basis of nearly degenerate constrained GHF solutions could also be used to parameterise spin Hamiltonians for $\tilde{S} > \frac{1}{2}$ systems in terms of zero-field splitting and *g*-tensors, thus opening an alternative simple and conceptually attractive way towards the calculation of a wider range of EPR parameters at a mean-field cost.

**Acknowledgement.** SGT thanks the German Academic Exchange Service (DAAD) for support in the early stages of this project.

## 5. Appendix

**A) Gradient-based optimisation including magnetisation constraints.** Parts of this section closely follow chapter 3.5 of Ref. [38]. Note that transformations between the original non-orthogonal atomic-orbital (AO) basis and an orthogonal AO (OAO) basis (we use Löwdin symmetric orthogonalisation) are necessary to avoid a costly transformation of two-electron integrals from the AO to the OAO basis. However, for ease of presentation, we assume that all integrals are given in the OAO basis. We trust that the reader can determine when a transformation from AO to OAO (or vice versa) is necessary in practice (see, e.g., Ref. [39]).

In Eq. (A1), a Thouless rotation $e^{\hat{Z}}$ relates $|\Phi\rangle$ to an initial guess $|\Phi^0\rangle$,

$$|\Phi\rangle = A e^{\hat{Z}} |\Phi^0\rangle ,  \quad (A1)$$

where

$$\hat{Z} = \sum_{v \in \text{virt}} \sum_{o \in \text{occ}} Z_{vo} \hat{c}_v^\dagger \hat{c}_o ,  \quad (A2)$$

and $A$ is a normalisation constant. The product $\hat{c}_v^\dagger \hat{c}_o$ of fermionic creation and annihilation operators transfers an electron from an occupied to a virtual orbital. Thouless parameters $Z_{vo}$ represent the unconstrained optimisation parameters and can be collected in a complex matrix $\mathbf{Z}$ of dimension $(M-N) \times N$, where $M$ is the total number of single-electron basis functions (twice the number of spatial basis functions), and $N$ is the number of electrons. The essence of GHF consists in minimising the single-determinant energy $E = \langle \Phi | \hat{H} | \Phi \rangle$, which requires an optimisation with respect to $\mathbf{Z}$.

An initial set of molecular orbitals (MOs) that define $|\Phi^0\rangle$ is provided in terms of OAO expansion coefficients in the columns of $\mathbf{O}^0_{\text{occ}}$ (an $M \times N$ matrix). The combined set of occupied and virtual MOs is a unitary $M \times M$ matrix, $\mathbf{O}^0 = (\mathbf{O}^0_{\text{occ}}, \mathbf{O}^0_{\text{virt}})$. In the first iteration, we set $\mathbf{Z} = \mathbf{0}$ and $\mathbf{O} = \mathbf{O}^0$. The GHF determinant $|\Phi\rangle$ is completely characterised (up to an irrelevant phase) by its density matrix $\boldsymbol{\rho}$,

$$\boldsymbol{\rho} = \mathbf{O}_{\text{occ}} \mathbf{O}_{\text{occ}}^\dagger .  \quad (A3)$$

In the GHF energy expression, Eq. (A4),

$$E = \langle \Phi | \hat{H} | \Phi \rangle = \frac{1}{2} \text{Tr}[\boldsymbol{\rho}(\mathbf{F} + \mathbf{h})] + V_{\text{nuc}} ,  \quad (A4)$$

**F** is the Fock matrix, **h** is the core Hamiltonian (which includes $\hat{H}_{SOC}$, Eq. (7)), and $V_{nuc}$ is the nuclear-repulsion energy (a constant).

In the following, we explain the calculation of the global energy gradient with respect to **Z**. Eqs. (A5)–(A10) describe the Thouless rotation (Eq. (A1)) of $\mathbf{O}^0$ into **O** in each optimisation step. Lower triangular matrices **L** ($N \times N$) and **M** [$(M-N) \times (M-N)$] are obtained by Cholesky decomposition, Eqs. (A5) and (A6):

$$\mathbf{1} + \mathbf{Z}^T \mathbf{Z}^* = \mathbf{L}\mathbf{L}^\dagger, \tag{A5}$$

$$\mathbf{1} + \mathbf{Z}^* \mathbf{Z}^T = \mathbf{M}\mathbf{M}^\dagger. \tag{A6}$$

Following Eqs. 3.45 and 3.47 in Ref. [38], we form the intermediate $\breve{\mathbf{O}}$,

$$(\breve{\mathbf{O}}_{occ})_o = (\mathbf{O}^0_{occ})_o + \sum_{v=1}^{M-N} Z_{vo} (\mathbf{O}^0_{virt})_v, \tag{A7}$$

$$(\breve{\mathbf{O}}_{virt})_v = (\mathbf{O}^0_{virt})_v - \sum_{o=1}^{N} Z^*_{vo} (\mathbf{O}^0_{occ})_o, \tag{A8}$$

where $(\breve{\mathbf{O}}_{occ})_o$ denotes the $o$-th column of $\breve{\mathbf{O}}_{occ}$, etc. Eqs. (A9) and (A10) represent the orthonormalisation of $\breve{\mathbf{O}}$ to yield a unitary **O**,

$$(\mathbf{O}_{occ})_{lm} = \sum_{k=1}^{N} (\mathbf{L}^{-1})_{lk} (\breve{\mathbf{O}}_{occ})_{km}, \tag{A9}$$

$$(\mathbf{O}_{virt})_{lm} = \sum_{k=1}^{M-N} (\mathbf{M}^{-1})_{lk} (\breve{\mathbf{O}}_{virt})_{km}. \tag{A10}$$

The global gradient vector **G** is defined in terms of variations of the GHF energy, Eq. (A11), caused by Thouless rotations from the reference state $|\Phi^0\rangle$.

$$\begin{aligned}\delta E &= \sum_{vo} \frac{\partial E}{\partial Z^*_{vo}} \delta Z^*_{vo} + \text{c.c.} = \\ &- \sum_{vo} \left( G_{vo} \delta Z^*_{vo} + \text{c.c.} \right) = \\ &-2 \sum_{vo} \left( \text{Re}(G_{vo}) \text{Re}(\delta Z_{vo}) + \text{Im}(G_{vo}) \text{Im}(\delta Z_{vo}) \right)\end{aligned} \tag{A11}$$

On the other hand, the local gradient $\mathbf{G}_{loc}$ is defined with respect to Thouless rotations from the present determinant $|\Phi\rangle$ (which determines **ρ** and **F**); its matrix form is given in Eq. (A12):

$$\mathbf{G}_{\text{loc}} = (\boldsymbol{\rho} - \mathbf{1})\mathbf{F}\boldsymbol{\rho} \ . \tag{A12}$$

The transformation of $\mathbf{G}_{\text{loc}}$ (computed in the OAO basis) to the MO basis of the global reference $|\Phi^0\rangle$ yields the global gradient, Eq. (A13):

$$\mathbf{G} = (\mathbf{O}_{\text{occ}}^0)^\dagger \mathbf{G}_{\text{loc}} \mathbf{O}_{\text{occ}}^0 \ . \tag{A13}$$

To reduce the number of iterations needed to reach convergence, we found it very useful to rescale the global gradient by dividing each element $G_{vo}$ by the square root of the corresponding virtual-occupied orbital-energy difference.

To constrain the magnetic moment to point along a general direction $\mathbf{n}$, we construct an orthonormal system of three-dimensional unit vectors, $\{\mathbf{n}, \mathbf{n}', \mathbf{n}''\}$, and demand that

$$\langle \Phi | \mathbf{n}' \cdot \hat{\boldsymbol{\mu}} | \Phi \rangle = \text{Tr}[\boldsymbol{\rho}(\mathbf{n}' \cdot \hat{\boldsymbol{\mu}})] = 0 \ , \tag{A14}$$

$$\langle \Phi | \mathbf{n}'' \cdot \hat{\boldsymbol{\mu}} | \Phi \rangle = \text{Tr}[\boldsymbol{\rho}(\mathbf{n}'' \cdot \hat{\boldsymbol{\mu}})] = 0 \ . \tag{A15}$$

(To instead constrain the spin direction, replace $\hat{\boldsymbol{\mu}}$ by $\hat{\mathbf{S}}$.) The local gradients for the two constrained quantities are obtained by replacing $\mathbf{F}$ in Eq. (A12) by $\mathbf{n}' \cdot \hat{\boldsymbol{\mu}}$ or $\mathbf{n}'' \cdot \hat{\boldsymbol{\mu}}$. The global gradients are again obtained by transforming to the MO basis of the global reference (cf. Eq. (A13)). To speed up convergence, the global constraint gradients are also divided by the square root of the orbital-energy difference.

The GHF energy $E$ and the global energy gradient $\mathbf{G}$, as well as the values of the two constrained quantities and their respective global gradients, and an initial MO set $\mathbf{O}^0$ (occupied and virtual orbitals) are passed to the `fmincon` function in `Matlab`. The separation into real and imaginary parts (Eq. (A11)) is required for `fmincon`, which optimises with respect to a set of real-valued variables.

**B) Connection between 3SCF and the Kramers-pair formalism for $E_{1/2}$ doublets in $C_{2v}^*$.**
As stated in the Results section, for all molecules in our test set (excluding $CH_4^+$ and $CuF_4^{2-}$), one of the three principal components of the $g$-tensor calculated for a single Kramers pair $\{|\Phi_\mathbf{n}\rangle, |\bar{\Phi}_\mathbf{n}\rangle\}$ based on Eq. (4), coincides with the 3SCF result, that is, $g_n = \langle \Phi_\mathbf{n} | \mathbf{n} \cdot \hat{\boldsymbol{\mu}} | \Phi_\mathbf{n} \rangle$ ($n =$ *x*, *y*, *z*). In the following, we rationalise this observation by tracing it to the fact that the cross-term of $\mathbf{n}\cdot\hat{\boldsymbol{\mu}}$ between Kramers partners vanishes, $\langle\Phi_\mathbf{n}|\mathbf{n}\cdot\hat{\boldsymbol{\mu}}|\bar{\Phi}_\mathbf{n}\rangle = 0$, due to symmetry.

All molecules in our set have either $C_{2v}$ symmetry or $C_{2v}$ is a subgroup of their molecular point group. From the three symmetry elements for each solution listed in Table 1, two arbitrarily chosen elements generate the self-consistent symmetry group $C_{2v}^*(C_2^*)$. Each of the three solutions has a distinct symmetry group, which is however in all cases isomorphic to $C_{2v}^*(C_2^*)$. For example, the (unitary) $C_2$ operation of the group $C_{2v}^*(C_2^*)$ corresponds to elements $\exp(-i\pi\hat{S}_x)\times\hat{\sigma}_{yz}$, $\exp(-i\pi\hat{S}_y)\times\hat{\sigma}_{xz}$ or $\exp(-i\pi\hat{S}_z)\times\hat{C}_2$ for $|\Phi_\mathbf{x}\rangle$, $|\Phi_\mathbf{y}\rangle$ or $|\Phi_\mathbf{z}\rangle$, respectively, where $\{\hat{E},\hat{C}_2,\hat{\sigma}_{xz},\hat{\sigma}_{yz}\}$ are the usual $C_{2v}$ elements of the nonrelativistic Hamiltonian $\hat{H}_0$. The $C_{2v}^*(C_2^*)$ symmetry implies that each pair $\{|\Phi_\mathbf{n}\rangle,|\bar{\Phi}_\mathbf{n}\rangle\}$ spans $E_{1/2}$ in the group $C_{2v}^*$. In other words, the Kramers pairs do not break the $C_{2v}^*$ symmetry of the Hamiltonian including SOC. (This is not true for arbitrary orientations $\mathbf{n}$, which do not yield GHF states with any self-consistent symmetry.) $E_{1/2}$ is the only fermionic irreducible representation in $C_{2v}^*$ [in a fermionic representation, a $2\pi$ spin rotation about an arbitrary axis introduces a factor of $(-1)$]. On the other hand, operators like $\hat{\mathbf{S}}$, $\hat{\mathbf{L}}$ or $\hat{\boldsymbol{\mu}}$ span bosonic representations. As an example, when dealing with $|\Phi_\mathbf{x}\rangle$, we attribute the Mulliken label $A_2$ to an operator that is symmetric (label A) with respect to $\exp(-i\pi\hat{S}_x)\times\hat{\sigma}_{yz}$ and antisymmetric (subscript 2) with respect to both $\exp(-i\pi\hat{S}_y)\times\hat{\sigma}_{xz}$ and $\exp(-i\pi\hat{S}_z)\times\hat{C}_2$.

Based on the foregoing symmetry analysis, we can now deduce that the cross-term of $\mathbf{n}\cdot\hat{\boldsymbol{\mu}}$ vanishes. For $|\Phi_\mathbf{n}\rangle$, $\mathbf{n}\cdot\hat{\boldsymbol{\mu}}$ transforms as $A_2$, while the other two (orthogonal) components of $\hat{\boldsymbol{\mu}}$ (that is, $\mathbf{n}'\cdot\hat{\boldsymbol{\mu}}$ and $\mathbf{n}''\cdot\hat{\boldsymbol{\mu}}$) transform as $B_1$ and $B_2$. As stated, $\{|\Phi_\mathbf{n}\rangle,|\bar{\Phi}_\mathbf{n}\rangle\}$ span $E_{1/2}$, but the two components of $E_{1/2}$ cannot be coupled by an $A_2$ operator (cf. Clebsch-Gordan coefficients for $C_{2v}^*$ given in Ref. [47]); thus, $\langle\Phi_\mathbf{n}|\mathbf{n}\cdot\hat{\boldsymbol{\mu}}|\bar{\Phi}_\mathbf{n}\rangle = 0$. In contrast, different components of $E_{1/2}$ can indeed be connected by $B_1$ or $B_2$ operators. Therefore, in general, $\langle\Phi_\mathbf{n}|\mathbf{n}'\cdot\hat{\boldsymbol{\mu}}|\bar{\Phi}_\mathbf{n}\rangle \neq 0$, if $\mathbf{n}^T\cdot\mathbf{n}' = 0$.

# 6. References

[1] L. F. Chibotaru, Adv. Chem. Phys. **153**, 397 (2013).
[2] A. Abragam and B. Bleaney, *Electron Paramagnetic Resonance of Transition Ions* (Dover, New York, 1986).
[3] P. J. Cherry, S. Komorovský, V. G. Malkin, and O. L. Malkina, Mol. Phys. **115**, 75 (2017).
[4] E. R. Sayfutyarova and G. K.-L. Chan, J. Chem. Phys. **148**, 184103 (2018).
[5] S. K. Singh, M. Atanasov, and F. Neese, J. Chem. Theory Comput. **14**, 4662 (2018).
[6] M. Kaupp, R. Reviakine, O. L. Malkina, A. Arbuznikov, B. Schimmelpfennig, and V. G. Malkin, J. Comput. Chem. **23**, 794 (2002).
[7] F. Neese, J. Chem. Phys. **122**, 34107 (2005).
[8] J. Gauss, M. Kállay, and F. Neese, J. Phys. Chem. A **113**, 11541 (2009).
[9] H. Bolvin, Chem. Phys. Phys. Chem. **7**, 1575 (2006).
[10] L. F. Chibotaru and L. Ungur, J. Chem. Phys. **137**, 64112 (2012).
[11] G. H. Lushington and F. Grein, J. Chem. Phys. **106**, 3292 (1997).
[12] S. Brownridge, F. Grein, J. Tatchen, M. Kleinschmidt, and C. M. Marian, J. Chem. Phys. **118**, 9552 (2003).
[13] D. Ganyushin and F. Neese, J. Chem. Phys. **138**, 104113 (2013).
[14] S. Vancoillie, P.-Å. Malmqvist, and K. Pierloot, ChemPhysChem **8**, 1803 (2007).
[15] M. Roemelt, J. Chem. Phys. **143**, 44112 (2015).
[16] P. Hrobárik, O. L. Malkina, V. G. Malkin, and M. Kaupp, Chem. Phys. **356**, 229 (2009).
[17] M. Repiský, S. Komorovský, E. Malkin, O. L. Malkina, and V. G. Malkin, Chem. Phys. Lett. **488**, 94 (2010).
[18] S. Gohr, P. Hrobarik, M. Repiský, S. Komorovský, K. Ruud, and M. Kaupp, J. Phys. Chem. A **119**, 12892 (2015).
[19] P. Verma and J. Autschbach, J. Chem. Theory Comput. **9**, 1052 (2013).
[20] D. Jayatilaka, J. Chem. Phys. **108**, 7587 (1998).
[21] P. J. Cherry, V. G. Malkin, O. L. Malkina, and J. R. Asher, J. Chem. Phys. **145**, 174108 (2016).
[22] J. Liu, H.-J. Koo, H. Xiang, R. K. Kremer, and M.-H. Whangbo, J. Chem. Phys. **141**, 124113 (2014).
[23] L. F. Chibotaru and L. Ungur, Phys. Rev. Lett. **109**, 246403 (2012).
[24] M. Gerloch and R. F. McMeeking, J. Chem. Soc. Dalt. Trans. **22**, 2443 (1975).
[25] C. Bloch, Nucl. Phys. **6**, 329 (1958).
[26] J. des Cloizeaux, Nucl. Phys. **20**, 321 (1960).
[27] S. Ghassemi Tabrizi, A. V Arbuznikov, and M. Kaupp, Chem. Eur. J. **24**, 4689 (2018).
[28] L. F. Chibotaru, A. Ceulemans, and H. Bolvin, Phys. Rev. Lett. **101**, 33003 (2008).
[29] I. Malkin, O. L. Malkina, V. G. Malkin, and M. Kaupp, J. Chem. Phys. **123**, 244103 (2005).
[30] J. Autschbach and B. Pritchard, Theor. Chem. Acc. **129**, 453 (2011).
[31] C. A. Jiménez-Hoyos, T. M. Henderson, and G. E. Scuseria, J. Chem. Theory Comput. **7**, 2667 (2011).
[32] S. Ghassemi Tabrizi, A. V Arbuznikov, and M. Kaupp, J. Phys. Chem. A **123**, 2361 (2019).
[33] D. Ganyushin and F. Neese, J. Chem. Phys. **125**, 24103 (2006).
[34] B. A. Hess, C. M. Marian, U. Wahlgren, and O. Gropen, Chem. Phys. Lett. **251**, 365 (1996).
[35] S. Koseki, M. W. Schmidt, and M. S. Gordon, J. Phys. Chem. **96**, 10768 (1992).
[36] E. van Lenthe, P. E. S. Wormer, and A. D. van der Avoird, J. Chem. Phys. **107**, 2488 (1997).

[37] P.-O. Löwdin, Phys. Rev. **97**, 1490 (1955).
[38] C. A. Jiménez-Hoyos, Ph.D. Thesis, Rice University, Houston, TX, 2013.
[39] A. Szabo and N. S. Ostlund, *Modern Quantum Chemistry* (Dover Publications, Courier Corporation, 1996).
[40] Gaussian 16, Revision C.01, M. J. Frisch, G. W. Trucks, H. B. Schlegel, G. E. Scuseria, M. A. Robb, J. R. Cheeseman, G. Scalmani, V. Barone, G. A. Petersson, H. Nakatsuji, X. Li, M. Caricato, A. V. Marenich, J. Bloino, B. G. Janesko, R. Gomperts, B. Mennucci, H. P. Hratchian, J. V. Ortiz, A. F. Izmaylov, J. L. Sonnenberg, D. Williams-Young, F. Ding, F. Lipparini, F. Egidi, J. Goings, B. Peng, A. Petrone, T. Henderson, D. Ranasinghe, V. G. Zakrzewski, J. Gao, N. Rega, G. Zheng, W. Liang, M. Hada, M. Ehara, K. Toyota, R. Fukuda, J. Hasegawa, M. Ishida, T. Nakajima, Y. Honda, O. Kitao, H. Nakai, T. Vreven, K. Throssell, J. A. Montgomery, Jr., J. E. Peralta, F. Ogliaro, M. J. Bearpark, J. J. Heyd, E. N. Brothers, K. N. Kudin, V. N. Staroverov, T. A. Keith, R. Kobayashi, J. Normand, K. Raghavachari, A. P. Rendell, J. C. Burant, S. S. Iyengar, J. Tomasi, M. Cossi, J. M. Millam, M. Klene, C. Adamo, R. Cammi, J. W. Ochterski, R. L. Martin, K. Morokuma, O. Farkas, J. B. Foresman, and D. J. Fox, Gaussian, Inc., Wallingford CT, 2016.
[41] R. A. Kendall, T. H. Dunning Jr, and R. J. Harrison, J. Chem. Phys. **96**, 6796 (1992).
[42] D. E. Woon and T. H. Dunning Jr, J. Chem. Phys. **98**, 1358 (1993).
[43] H. Fukutome, Int. J. Quantum Chem. **20**, 955 (1981).
[44] M. Ozaki and H. Fukutome, Prog. Theor. Phys. **60**, 1322 (1978).
[45] J.-P. Blaizot and G. Ripka, *Quantum Theory of Finite Systems* (The MIT Press, Cambridge, MA, 1985).
[46] M. Tinkham, *Group Theory and Quantum Mechanics* (McGraw-Hill, New York, 1964).
[47] S. L. Altmann and P. Herzig, *Point-Group Theory Tables* (Clarendon Press, Oxford, 1994).
[48] C. van Wüllen, J. Chem. Phys. **130**, 194109 (2009).
[49] M. R. Pederson and S. N. Khanna, Phys. Rev. B **60**, 9566 (1999).
[50] R. Reviakine, A. V Arbuznikov, J. C. Tremblay, C. Remenyi, O. L. Malkina, V. G. Malkin, and M. Kaupp, J. Chem. Phys. **125**, (2006).
[51] R. Seeger and J. A. Pople, J. Chem. Phys. **66**, 3045 (1977).
[52] B. C. Huynh and A. J. W. Thom, J. Chem. Theory Comput. **16**, 904 (2019).
[53] V. I. Lebedev and D. N. Laikov, Dokl. Math. **59**, 477 (1999).
[54] C. A. Jiménez-Hoyos, T. M. Henderson, T. Tsuchimochi, and G. E. Scuseria, J. Chem. Phys. **136**, 164109 (2012).
[55] J. K. Percus and A. Rotenberg, J. Math. Phys. **3**, 928 (1962).
[56] N. Lee and A. J. W. Thom, J. Chem. Theory Comput. **18**, 710 (2022).
[57] P. J. Lestrange, D. B. Williams-Young, A. Petrone, C. A. Jiménez-Hoyos, and X. Li, J. Chem. Theory Comput. **14**, 588 (2018).
[58] M. Gräf and D. Potts, Numer. Funct. Anal. Optim. **30**, 665 (2009).
[59] S. Mamone, G. Pileio, and M. H. Levitt, Symmetry. **2**, 1423 (2010).
[60] J. S. Griffith, *The Theory of Transition-Metal Ions* (Cambridge University Press, 1964).
[61] A. Schäfer, H. Horn, and R. Ahlrichs, J. Chem. Phys. **97**, 2571 (1992).
[62] F. Weigend and R. Ahlrichs, Phys. Chem. Chem. Phys. **7**, 3297 (2005).
[63] E. Papajak, J. Zheng, X. Xu, H. R. Leverentz, and D. G. Truhlar, J. Chem. Theory Comput. **7**, 3027 (2011).